%% file: bare_jrnl_compsoc.tex
\documentclass[10pt,journal,compsoc]{IEEEtran}
\ifCLASSOPTIONcompsoc
  \usepackage[nocompress]{cite}
\else
  \usepackage{cite}
\fi
\ifCLASSINFOpdf
\else
\fi

\usepackage{url}
\usepackage{multirow}
\usepackage{bm}
\usepackage{lipsum}
\usepackage{enumitem}
\usepackage{xspace}
\usepackage{subfig}
\usepackage{amsmath}
\usepackage{color}
\usepackage{graphicx}
\usepackage{booktabs}
\usepackage{hyperref}
\usepackage{amssymb}
\usepackage[linesnumbered,boxed,ruled,vlined,noend]{algorithm2e}

\usepackage[labelfont=bf, textfont=md, font=small,aboveskip=0pt,belowskip=0pt]{caption}

\usepackage{amsthm}
\newtheorem{theorem}{Theorem}

\newcommand{\diff}[1]{{\color{black}{#1}}}
\newcommand{\revise}[1]{{\color{black}{#1}}}
\newcommand{\vpara}[1]{\vspace{0.01in}\noindent\textbf{#1.}}

\newcommand{\jz}[1]{\textcolor{red}{[jiezhong:*#1*]}}

\newcommand{\hide}[1]{} 
\newcommand{\beq}[1]{{\small\begin{equation}#1\end{equation}}}

\newcommand{\besp}[1]{\begin{split}#1\end{split}}

\DeclareMathOperator{\tln}{trunc\_log}
\DeclareMathOperator{\htln}{trunc\_log^\circ}
\DeclareMathOperator{\vol}{vol}
\input{math_commands.tex}

\newcommand{\lightne}{\textsc{LightNE}\xspace}

\newcommand{\lightneplus}{\textsc{LightNE\,2.0}\xspace}
\def\gvt{GraphVite}
\def\pbg{PyTorch-BigGraph}
\def\shortpbg{PBG}

\newcommand{\myparagraph}[1]{\vspace{1.5pt}\noindent {\bf #1.}}

\begin{document}
%
\title{
\diff{Towards Lightweight and Automated Representation Learning System for Networks}
}
%
%
%
%

\author{Yuyang~Xie, 
        Jiezhong~Qiu, 
        Laxman~Dhulipala,
        Wenjian~Yu,
        Jie~Tang,
        Richard~Peng,
        and Chi~Wang
\IEEEcompsocitemizethanks{\IEEEcompsocthanksitem Y. Xie, J. Qiu, W. Yu and J. Tang are with Tsinghua University. Email: \{xyy18, qiujz16\}@mails.tsinghua.edu.cn, \{yu-wj, 
jietang\}@tsinghua.edu.cn.
L. Dhulipala is with University of Maryland. Email: laxman@umd.edu. 
R. Peng is with University of Waterloo. Email: 
y5peng@uwaterloo.ca. C. Wang is with Microsoft Research. Email: wang.chi@microsoft.com.
(Y. Xie and J. Qiu contributed equally).
} 

\thanks{Manuscript received 29 Apr. 2022; revised 3 Oct. 2022; accepted 27 Jan. 2023.
This work was supported by NSFC under Grant 61872206,  the National Key R\&D Program of China (2018YFB1402600),
NSFC for Distinguished Young Scholar~(61825602), NSFC~(61836013),
and in part by an NSERC Discovery Grant. (Corresponding authors: W. Yu, J. Tang and C. Wang.)
}
}

%
%

\markboth{Journal of \LaTeX\ Class Files,~Vol.~14, No.~8, August~2015}%
{Shell \MakeLowercase{\textit{et al.}}: Bare Demo of IEEEtran.cls for Computer Society Journals}
%




\input{0.1.abstract.tex}

\maketitle
\IEEEdisplaynontitleabstractindextext

%
\IEEEpeerreviewmaketitle

\input{1.intro.tex}
\input{2.related.tex}
\input{3.back_and_ne.tex}

\begin{figure*}[!t]
  \centering
  \includegraphics[width=0.9\textwidth]{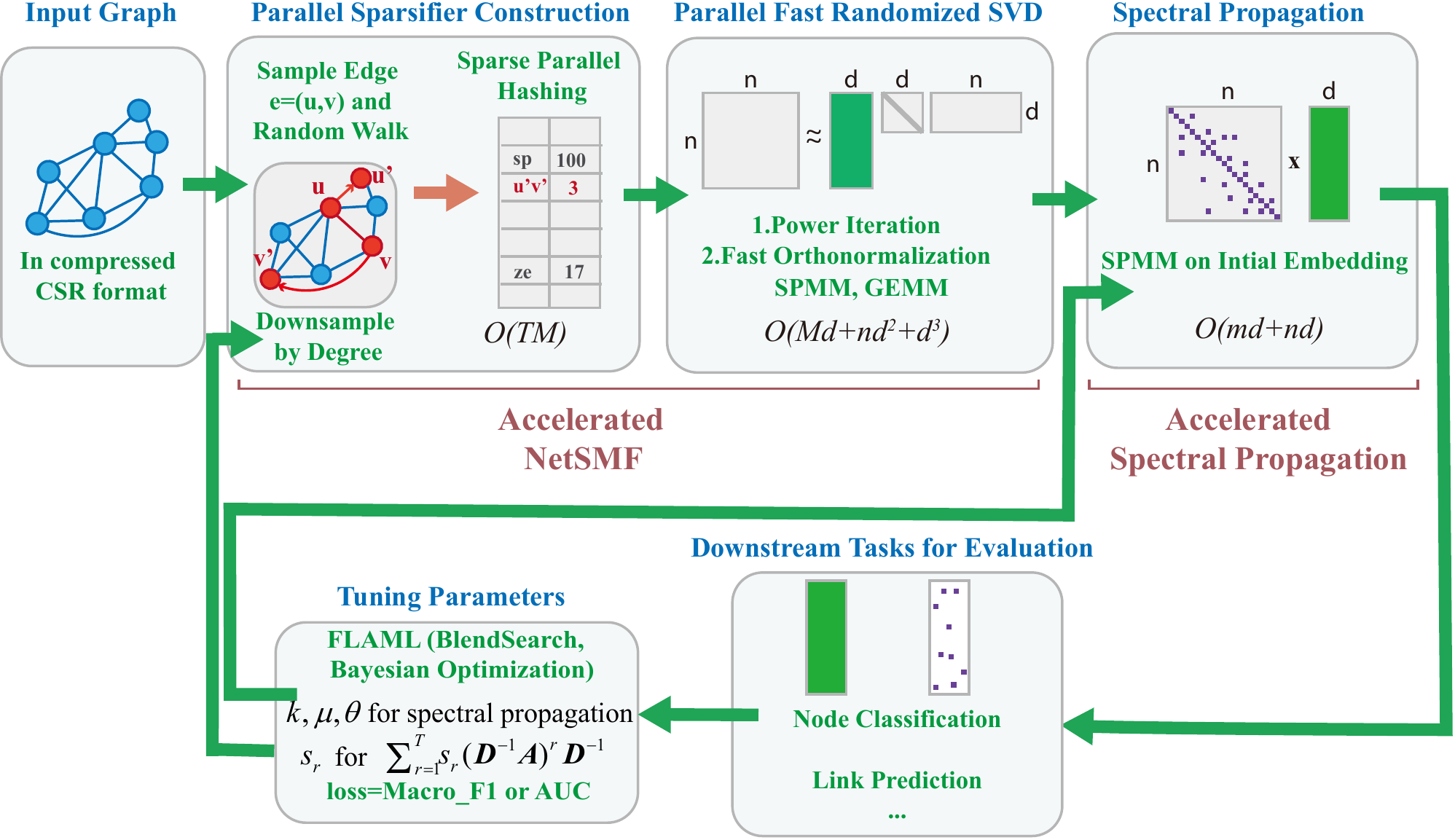}
  \caption{\diff{System overview of \lightneplus{}.}}
  \label{fig:system_overview}
  \vspace{-0.2in}
\end{figure*}

\input{4.system.tex}

\input{5.eval.tex}

\input{6.conclusion.tex}
\ifCLASSOPTIONcaptionsoff
  \newpage
\fi



\bibliographystyle{IEEEtran}
\bibliography{IEEEabrv}
\end{document}

%% file: math_commands.tex
\usepackage{amsmath,amsfonts,bm}

\def\Figref#1{Figure~\ref{#1}}

\def\Secref#1{Section~\ref{#1}}

\def\Eqref#1{Equation~\eqref{#1}}

\def\1{\bm{1}}

\def\mA{{\bm{A}}}

\def\mD{{\bm{D}}}

\def\mI{{\bm{I}}}

\def\mL{{\bm{L}}}
\def\mM{{\bm{M}}}

\def\mU{{\bm{U}}}
\def\mV{{\bm{V}}}

\def\mX{{\bm{X}}}

\DeclareMathAlphabet{\mathsfit}{\encodingdefault}{\sfdefault}{m}{sl}
\SetMathAlphabet{\mathsfit}{bold}{\encodingdefault}{\sfdefault}{bx}{n}

\newcommand{\E}{\mathbb{E}}



%% file: 0.1.abstract.tex
\IEEEtitleabstractindextext{%
\begin{abstract}
    We propose \lightneplus{},
    a cost-effective, scalable, automated, and high-quality network embedding system that scales to graphs with hundreds of billions of edges on a single machine. In contrast to the mainstream belief that distributed architecture and GPUs are needed for large-scale network embedding with good quality, we prove that we can achieve higher quality, better scalability, lower cost, and faster runtime with shared-memory, CPU-only architecture. 
    \lightneplus{} combines two theoretically grounded embedding methods NetSMF and ProNE.
    We introduce the following techniques to network embedding for the first time: 
    (1) a newly proposed downsampling method to reduce the sample complexity of NetSMF while preserving its theoretical advantages;
    (2) a high-performance parallel graph processing stack GBBS to achieve high memory efficiency and scalability; 
    (3)  sparse parallel hash table to aggregate and maintain the matrix sparsifier in memory;
    (4) a fast randomized singular value decomposition~(SVD) enhanced by power iteration and fast orthonormalization to improve vanilla randomized SVD in terms of both efficiency and effectiveness;
    (5) Intel MKL for proposed fast randomized SVD and spectral propagation;
    and (6) a fast and lightweight AutoML library FLAML 
    for automated hyperparameter tuning. 
    \revise{Experimental results show that \lightneplus{} can be up to 84\(\times\) faster than \gvt{}, 30\(\times\) faster than \shortpbg{} and 9\(\times\) faster than NetSMF while delivering better performance. \lightneplus{} can embed very large graph with 1.7 billion nodes and 124 billion edges in half an hour on a CPU server, while other baselines cannot handle very large graphs of this scale.}
\end{abstract}

}

%% file: 1.intro.tex
\IEEEraisesectionheading{\section{Introduction}\label{sec:introduction}}

%
%
%
%
\IEEEPARstart{E}{-commerce} and social networking companies today face the challenge of analyzing and mining graphs with billions of nodes, and tens of billions to trillions of edges.
In recent years, a popular learning approach has been to apply network embedding techniques to obtain a vector representation of each node.
These learned representations, or embeddings, can be easily consumed in downstream machine learning and recommendation algorithms.
These representations are widely used in various online services and are updated frequently~\cite{ying2018graph, wang2018billion, ramanath2018towards}.
For example, one of the core item-recommendation systems at Alibaba with billions of items and users
requires frequent re-embedding as both new users and items arrive online, and the underlying embedding must be quickly recomputed~\cite{wang2018billion}.
A similar system at LinkedIn computes embeddings of millions of individuals (nodes) offline and must periodically re-embed this graph  to maintain high accuracy~\cite{ramanath2018towards}.
In both scenarios, computing embeddings must be done scalably
and with low latency.

Despite a significant amount of research on developing sophisticated network embedding algorithms~\cite{tang2015line, perozzi14deepwalk, grover2016node2vec},
using simple and scalable embedding solutions that potentially sacrifice a significant amount of accuracy remains the primary choice in the industry for dealing with large-scale graphs.
For example, LinkedIn uses LINE~\cite{tang2015line} for embedding, which only captures local structural information within nodes' 1-hop  neighborhoods.
Alibaba embeds a 600-billion-node commodity graph by first partitioning it into 12,000 50-million-node subgraphs, and then embedding each subgraph separately with 100 GPUs running DeepWalk~\cite{perozzi14deepwalk}.
The reason is that in practice, graphs are updated frequently, and the embedding algorithms are often
required to run every few hours~\cite{wang2018billion}.

While many new embedding systems have been proposed in the literature that demonstrate high accuracy for downstream applications,
the high latency, limited scalability, and high computational cost prohibit these techniques from large-scale deployment or commercial usage on massive datasets.
For example, GraphVite~\cite{zhu2019graphvite} is a CPU-GPU hybrid system based on DeepWalk, which takes 20 hours to train on the Friendster graph~(65M nodes and 1.8B edges) with 4 P100 GPUs.
The cost of GraphVite for obtaining the embedding on this graph is 210 dollars measured by cloud virtual machine rent. One can estimate that embedding 10,000 such graphs~(following the Alibaba approach) using GraphVite would amount to over 2 million dollars per run, which is prohibitively costly. 


Motivated by the desire to obtain accurate, highly scalable, and cost-effective solutions that can embed networks with billions of nodes and hundreds of billions of edges, \diff{we design \lightneplus{}\footnote{Our code is available at \url{https://github.com/xptree/LightNE/tree/frsvds}.}. Our design has the following objectives:}
\begin{enumerate}[label=(\textbf{\arabic*}),topsep=1pt,itemsep=0pt,parsep=0pt,leftmargin=20pt]
    \setlength\itemsep{0pt}
    \diff{\item {\bf Scalable}: Embed graphs with 124B edges within half an hours.}
    \item {\bf Lightweight}: Occupy hardware costs below 100 dollars measured by cloud rent to process 1B to 100B edges.
    \item {\bf Accurate}: Achieve the highest accuracy in downstream tasks
          under the same time budget and similar resources.
    \diff{\item {\bf Automated}: Fast and automated hyperparameter tuning to further boost embedding quality.}
\end{enumerate}

\myparagraph{Our Techniques}
To reduce both cost and latency, we use a single-machine shared-memory environment equipped with multi-core CPUs, which are ubiquitous from cloud-providers today.
Furthermore, to optimize our processing times and fully utilize the system, we avoid using SSDs or other external storage and instead utilize enough RAM so that both the input graph and all the intermediate steps can fit into memory, e.g., 1.5TB of RAM.
Purchasing or renting a system with sufficient RAM and multi-core CPU(s) is the dominant cost of our system.
Even with such a simple architecture, we successfully meet all our design goals by leveraging the following techniques and building an integrated system:


Firstly, we combine two lines of advances in efficient and effective network embedding techniques, \textbf{sample-based approximation} and \textbf{spectral approximation} of random walks stemming from the original DeepWalk: NetSMF~\cite{qiu2019netsmf} and ProNE~\cite{zhang2019prone}.
Instead of using the general-purpose and computationally
inefficient stochastic gradient descent method from most other solutions,
both NetSMF and ProNE perform principled, cheap matrix operations on graphs to leverage the unique characteristics of real-world graphs such as sparsity, power-law degree distribution, and spectral properties. We combine both sample-based approximation and spectral approximation of random walks to achieve high accuracy while maintaining their advantages of low resource consumption and efficiency on real-world graphs.

Secondly, we propose  \textbf{a new edge sampling algorithm} that reduces the number of required random-walk samples of original NetSMF by a factor of \#edges/\#vertices. On real-world graphs, it achieves a $10$-$100\times$ reduction in the number of samples. Our algorithm is grounded in \textbf{spectral graph sparsification theory}.

\diff{
Thirdly, we propose a \textbf{fast randomized SVD algorithm} for the sparse matrix generated after spectral graph sparsification. The proposed \textbf{fast randomized SVD algorithm} is improved by combining power iteration scheme and fast orthonormalization operation with vanilla randomized SVD algorithm used in ~\cite{qiu2021lightne}.
}

Fourthly, we optimize our system for commodity shared-memory architectures and performing sparse matrix operations by
(1) utilizing state-of-the-art \textbf{shared-memory graph processing techniques}, including parallel graph compression and efficient bulk-parallel operations,
(2) integrating efficient parallel data structures and techniques such as \textbf{sparse hash tables} for random walk samplers, and
(3) using Intel Math Kernel Library~(MKL) for linear algebra operations.
These techniques enable us to achieve between 4--32$\times$ speedup over state-of-the-art network embedding systems, while also experiencing a similar order of magnitude cost improvement, all while maintaining or improving accuracy. Our memory efficiency enables us to scale to graphs significantly larger than those processed by single-machine embedding systems today. \diff{In particular, we show that using \lightneplus{} we can embed one of the largest publicly available graphs, the WebDataCommons hyperlink 2014 graph, with over 100 billion edges in half an hour.}
\diff{
Lastly, we propose to search the hyperparameters of \lightneplus{} with \textbf{automated machine learning (AutoML)}. We leverage a fast and lightweight AutoML library, FLAML~\cite{wang2021flaml}. Compared to the default hyperparameters, the searched ones bring significant performance improvement in various datasets.
}

\revise{In experiment, we use three groups of datasets to evaluate \lightneplus{}. The small graphs are used to verify the effectiveness of \lightneplus{}, compared with all the baselines. Then we test \lightneplus{} on large graphs which are used in previous works, such as Livejournal studied by \pbg{}~(\shortpbg{}), Friendster studied by \gvt{} and OAG studied by NetSMF. The very large graphs are used to demonstrate that our system can scale better than previous work.}
Compared to three large-scale systems: \gvt{}, \shortpbg{}, and NetSMF, and using the tasks and the largest datasets evaluated by each system 
\lightneplus{} takes an order of magnitude lower latency and cost, while achieving the state-of-the-art accuracy.
Compared to ProNE, our accuracy is significantly higher while the latency is comparable.
In addition, we show that our system can scale to networks with billions of nodes, and hundreds of billions of edges on a single machine, which has never been demonstrated by any existing network embedding systems, including ProNE.

\diff{
This article is an extension of prior work~\cite{qiu2021lightne} which proposes \lightne{}.
Compared to the prior work, we have the following new contributions: 
(1) proposal of a new randomized SVD algorithm via fast orthonormalization and power iteration (\Secref{subsec:frsvd}),
which is not only faster but also more accurate than the vanilla randomized SVD in \cite{qiu2021lightne};
(2) tuning hyperparameters with AutoML library FLAML~\cite{wang2021flaml} to further improve the overall performance (\Secref{sec:auto}); 
(3) incorporating the above techniques into \lightne{} and developing \lightneplus{}, which is a lightweight and automated network embedding system (\Figref{fig:system_overview});
(4) extensive experiments on real-world graphs, detailed comparison between \lightneplus{} and existing baselines including \lightne{} (\Secref{sec:eval}); 
}




%% file: 2.related.tex
\vspace{-0.1in}
\section{Related Work}
\label{sec:related}

We review related work
of network embedding algorithms/systems.

\hide{
    \begin{table}[htbp]
        \caption{Comparison of marjor Network Embedding System.}
        \label{tbl:comparison}
        \begin{center}
            \small
            \begin{tabular}{l l l l}
                \toprule
                System      & Optimization & Data Storage & Accelorator    \\
                \midrule
                \method     & explict MF   & CSR          & CPU            \\
                NetSMF      & explict MF   & COO          & CPU            \\
                \gvt{}      & gradient     & COO          & CPU-GPU hybrid \\
                \shortpbg{} & gradient     & COO          & CPU            \\
                \bottomrule
            \end{tabular}
            \normalsize
        \end{center}
    \end{table}
}

\vpara{Network Embedding Algorithms}
Over the last decade,
network embedding algorithms have been extensively studied.
A survey can be found in~\cite{hamilton2017representation}.
From an optimization aspect,
recent network embedding algorithms fall into
three main categories.
The first category uses general-purpose stochastic gradient
descent to optimize a logistic loss and follows the skip-gram model framework~\cite{mikolov2013distributed}.
Methods belonging to this category
include DeepWalk~\cite{perozzi14deepwalk}, LINE~\cite{tang2015line}, and node2vec~\cite{grover2016node2vec}.
To date, the only bounds on sample efficiency and convergence rate for
these methods require additional assumptions~\cite{de2014global}.
The second category uses singular value decomposition~(SVD) or other matrix approximation techniques to obtain the best low-rank approximations~\cite{eckart1936approximation}.
Examples of methods in this category include GraRep~\cite{cao2015grarep},
HOPE~\cite{ou2016asymmetric}, NetMF~\cite{qiu2018network},
NetSMF~\cite{qiu2019netsmf}, RandNE\cite{zhang2018billion}, FastRP\cite{chen2019fast}, and ProNE~\cite{zhang2019prone}.
\lightneplus{} also belongs to this category.
Graph Neural Networks represent the third line of
network embedding algorithms~\cite{battaglia2018relational}.
Such methods include GCN~\cite{kipf2016semi}, GAT~\cite{velivckovic2017graph}, GIN~\cite{xu2018powerful}, GraphSAGE~\cite{hamilton2017representation}
and PinSAGE~\cite{ying2018graph}.
These algorithms usually rely on vertex attributes,
as well as supervised information.
They are beyond the scope of this paper because our focus is on
graphs with no additional information.

\vpara{Network Embedding Systems}
Due to the efficiency challenges posed by large
graphs, several systems for embedding large graphs have been developed.
We give a brief overview of the  most
related
and
comparable ones.
\textit{\gvt{}}~\cite{zhu2019graphvite} is a CPU-GPU hybrid network embedding system based on DeepWalk~\cite{perozzi14deepwalk} and LINE~\cite{tang2015line}.
It uses CPU to conduct graph operations and GPU to perform linear algebra operations.
The system is bounded by GPU memory, which in most cases
is at most 12GB per GPU: embedding graphs with billions of vertices
often require hundreds of Gigabytes of parameter memory.
This limit constraints \gvt{} to repeatedly updating
only a small part of the embedding matrix.
\textit{\pbg{}}~\cite{lerer2019pytorch} is a distributed memory
system based on DeepWalk~\cite{perozzi14deepwalk} and LINE~\cite{tang2015line}.
It uses graph partition for load balancing
and a shared parameter server for synchronization.
\lightneplus{} is designed for shared memory machines, 
where communication
is much cheaper than distributed memory systems.
\textit{NetSMF}~\cite{qiu2019netsmf} is a network embedding system based on
sparse matrix factorization.
The system is built on OpenMP
and Eigen3~(a C++ template library for linear algebra).
On large graphs, NetSMF is still time-consuming
due to its poor implementation of the graph processing system and the shortcoming of Eigen3 in supporting sparse matrix operations.
Our proposal contains a redesign of NetSMF
that focuses on the performances on graphs and sparse matrices.
A detailed experimental comparison of \lightneplus{} and NetSMF
is in \Secref{sec:eval}.
\textit{NRP}~\cite{yang2020homogeneous} is a recently proposed network embedding system built upon Matlab. It derives embeddings from the pairwise personalized PageRank~(PPR) matrix.
Although it is also based on random walks, it omits a step of taking the entry-wise logarithm of the random walk matrix before factorization, which is a required step by NetMF and NetSMF for establishing the equivalence to DeepWalk. Due to that omission, NRP is able to operate on the original graph efficiently while the others must construct the random walk matrix exactly or approximately.




%% file: 3.back_and_ne.tex
\section{Background}

\begin{table}[tb]
  \centering
  \caption{Notation used throughout this paper.}
  \label{tbl:notation}
  \footnotesize
  \begin{tabular}[htbp]{@{}l@{}|@{}r|l@{}|r@{}}
    \toprule
    \textbf{Notation} & \textbf{Description} & \textbf{Notation} & \textbf{Description}                    \\  \midrule
    $G$               & input network        & $b$               & \#negative samples                      \\
    $V$               & vertex set, $|V|=n$  & $T$               & context window size                     \\
    $E$               & edge set, $|E|=m$    & $\mX$             & $n\times d$ embedding matrix            \\
    $\bm{A}$          & adjacency matrix     & $k$               & spectral propagation steps              \\
    $\bm{D}$          & degree matrix        & $\mD-\mA$         & graph Laplaician      $\mL$             \\
    $\vol{(G)}$       & volume of $G$        & $\mI-\mD^{-1}\mA$ & normalized  Laplacian   $\mathcal{\mL}$ \\
    $M$               & \# edge samples      & $d$               & embedding dimension                     \\
    $q$               & power parameter      & $s$               & oversampling parameter                    \\
    \bottomrule
  \end{tabular}
\end{table}

\label{sec:ne}
We provide a self-contained background of the fundamental embedding techniques of our system. The list of notations used in this paper can be found in Table~\ref{tbl:notation}.
We take a matrix-oriented view of graph embedding: the resulting embedding vectors are simply the rows of a $n \times d$ matrix $\mX$ that we compute from the original adjacency matrix $\bm{A}$ of $n\times n$, where $n$ and $d$ denote the number of nodes and the embedding dimension, respectively.
This interpretation of embedding gives us the flexibility of utilizing multiple matrix processing tools and synthesizing them.




\vpara{NetMF}
We begin with the matrix factorization approach introduced in~\cite{qiu2018network}, which showed that most network embedding methods up to that point in time, including DeepWalk~\cite{perozzi14deepwalk} LINE~\cite{tang2015line}, and node2vec~\cite{grover2016node2vec}, can be described as factorizing a matrix polynomial of the adjacency matrix $\bm{A}$ and degree matrix $\bm{D}$ of the graph.
Formally, for an unweighted, undirected graph, $\bm{A}$ is the matrix with $1$ in every entry with an edge, and $0$ everywhere else; and the matrix $\bm{D}$ is the diagonal matrix where the $i$-th diagonal entry contains the degree of the $i$-th vertex.
The core result by \cite{qiu2018network} is that DeepWalk can be viewed as approximately factorizing the following matrix
\beq{
  \label{eq:NE}
  \diff{\mM \triangleq \htln\left(\frac{\vol(G)}{b}\sum_{r=1}^T s_r(\bm{D}^{-1}\bm{A})^r \bm{D}^{-1}\right)}
}where $T$ represents the length of sliding window (by default $T=10$), \diff{
$s_1 = \cdots = s_T = \frac{1}{T}$,}
$\htln$ is the truncated logarithm applied
entry-wise to a matrix
($\tln(x) = \max\{0, \log{x} \}$), and $\vol(G)=2m$
is the total number of edges in $G$.
Moreover, LINE approximately factorizes a matrix in the same form but for $T=1$.
The bottleneck of factorizing the matrix in \Eqref{eq:NE} is that  $(\bm{D}^{-1}\bm{A})^r$ tends to be a dense matrix as the increase of $r$,
and thus constructing the matrix is cost-prohibitive even before the factorization can be performed, due to the sheer amount of memory required. Note that the truncated logarithm is critical for embedding quality and cannot be omitted, otherwise there exists a shortcut to the factorization without constructing the dense matrix, similar to NRP~\cite{yang2020homogeneous}.

\begin{algorithm}[tb]
  \small
  \SetAlgoLined \DontPrintSemicolon
  \SetKwInOut{Input}{Input}
  \SetKwInOut{Output}{Output}
  \SetKwFunction{pathsample}{PathSample}
  \SetKwProg{myalg}{Procedure}{}{}
  \myalg{\pathsample{$G$, $u$, $v$, $r$}}{
    Let a random edge $(u,v)$ be given.\\
    Sample a random number $s$ uniformly in $[0, r - 1]$.\\
    $u' \gets $ random walk $u$ for $s$ steps on graph $G$\\
    $v' \gets $ random walk $v$ for $r - 1 - s$ steps on graph $G$.\\
    \KwRet edge $(u', v')$\\
  }{}
  \normalsize
  \caption{PathSampling.}
  \label{alg:pathsample}
\end{algorithm}

\vpara{NetSMF}
One approach to mitigate the increased construction cost for constructing the matrix in \Eqref{eq:NE} is through the sampling of random walks.
Qiu et al.~\cite{qiu2019netsmf} showed that an $r$-step random walk matrix $(\bm{D}^{-1}\bm{A})^r$ could be approximated by repeating the PathSampling algorithm~(Algo.~\ref{alg:pathsample}). 
The result of applying Algo.~\ref{alg:pathsample} is a $r$-step walk in $\bm{A}$ and contributes to a non-zero entry to a sparsified version of $(\bm{D}^{-1}\bm{A})^r$.
Building upon an analysis of sparsification of random walk matrix polynomials~\cite{cheng15sample}, Qiu et al.~\cite{qiu2019netsmf} showed that a nearly-linear number of samples  w.r.t. the number of edges in $G$~(i.e., $m$) is sufficient to make a spectral approximation of  $(\bm{D}^{-1}\bm{A})^r$.
They then demonstrated experimentally that this matrix could be used in place of the dense random walk matrix used by NetMF.
After constructing the sparse matrix, NetSMF factorizes it with vanilla randomized  SVD~\cite{HalkoMT11}.
Improving the scalability of NetSMF, especially the efficient sampling of random walks and fast randomized SVD, is the starting point of our system.

\vpara{ProNE}
ProNE~\cite{zhang2019prone} proposed to firstly conduct SVD on
a matrix $\bm{M}$ with each entry defined to be $\bm{M}_{uv} \triangleq
  \log{
    \left(
    \frac{\bm{A}_{uv}}{\bm{D}_u}
    \frac{\sum_{j} \left( \sum_{i} \mA_{ij}/\mD_i\right)^\alpha}{b \left(\sum_{i} \mA_{iv}/\mD_i\right)^\alpha}
    \right)
  }$, which is a modulated  normalized graph Laplacian, with  $b=1$ and $\alpha=0.75$ by default.
Given the factorized embedding matrix $\bm{X}$, ProNE applies
a filter to each column of the matrix using a low degree
polynomial in the normalized graph Laplacian matrix $\mathcal{\mL} \triangleq \mI - \mD^{-1}\mA$, i.e.,
$
  \sum_{r = 0}^{k} c_{r} \mathcal{\mL}^{r} \bm{X}
$, where $c_r$'s are chosen to be
coefficients of Chebyshev polynomials and $k$ is set to around $10$.
We will utilize the same choice of parameters for
spectral propagation step, but instead apply it to the  factorization of the sparsified NetMF matrix in \Eqref{eq:NE}.
\hide{
  Algorithmically, this transformation is highly efficient.
  It does not need to evaluate the higher powers of $\mathcal{\mL}$,
  but rather only applies repeated matrix-vector
  multiplications to the columns of $\bm{X}$.
  As $\bm{X}$ has only $d$ columns, this leads to a total cost
  of $O(mkd)$, significantly less than the cost of computing
  or approximating the random walk matrix polynomial directly, which could have $O(n^3)$ complexity.
}

\hide{
  The performance of such a propagation scheme critically
  depends on the choice of the $c_r$'s.
  In~\cite{zhang2019prone}, they were chosen to be
  coefficients of Chebyshev polynomials, and $k$ was set to around $10$.
  We will utilize a similar choice of parameters for this
  post-processing step, but instead apply it to the factorization
  of the~(approximate) random walk matrix described in
  Sections~\ref{subsec:NetMF}~and~\ref{subsec:NetSMF}.
}

\section{L\texorpdfstring{\MakeLowercase{\textsc{ight}}}{ight}NE 2.0: Algorithm Design}
\label{subsec:algorithm}
From an algorithm design perspective,
our design of \lightneplus{} combines NetSMF, ProNE and proposed fast randomized SVD.
In particular,
\lightneplus{} consists of three steps.
The first step is to incorporate a novel edge
downsampling algorithm into the sparisfier construction of 
NetSMF, which significantly improves sample complexity.
\diff{The second step is to obtain the initial embedding with proposed fast randomized SVD algorithm.}
The third step is to enhance the initial embedding using ProNE's spectral propagation. We then introduce the three steps in detail.

\subsection{Sparsifier Construction with Edge Downsampling}
\label{subsec:algo_sparsifier}
NetSMF~\cite{qiu2019netsmf} has proposed an efficient PathSampling algorithm
to approximate the
$r$-step random walk matrix, $(\mD^{-1}\mA)^r$, with roughly $O(m)$ samples.
However, for graphs with billions of edges, there is still an urgent need
to further reduce its sample complexity while preserving its theoretical advantages.
Our approach is to downsample edges that will be added to the sparsifier.
We do this by adding a further layer of sampling to Algo.~\ref{alg:pathsample} --- for each sampled edge $e=(u, v)$,
we flip a coin that comes up heads with some  probability $p_e$, then only apply Algo.~\ref{alg:pathsample}
and add the sampled edge to the sparsifier with adjusted weight $A_{u,v}/p_e$ if the coin comes up heads.
Such a sampling method is a special case of importance sampling ---
adjusting edge weights
ensures
the downsampled graph is an unbiased estimation to the original graph in terms of the graph Laplacian:
\begin{theorem}[Unbiasness of Edge Downsampling, Sec 6.5 in \cite{teng2016scalable}]
  Let the graph Laplacian of the original graph be $\mL_{G} \triangleq \mD -\mA$. Note that
  $\mL_G=\sum_{(u,v)\in E}A_{u,v} \mL_{u,v}$ where $\mL_{u, v}$ is the Laplacian matrix of the graph with just one unweighted edge between $u$ and $v$. Also denote the downsampled graph to be $H$, then we have
  $
    \E[\mL_H] = \sum_{e=(u,v)\in E} p_e \frac{A_{u, v}}{p_e} \mL_{u, v} = \mL_G
  $.
\end{theorem}
\noindent In theory, setting sampling probability $p_e$ as an upper bound of the
effective resistance~\cite{spielman2011graph}  guarantees  an accurate approximation of input graph $G$ with high probability, i.e., $p_e \leftarrow \min(1, C A_{u,v} R_{u,v})$ where  $R_{u,v}$ is the effective resistance between $u$ and $v$, and $C$ is some constant.
However,
how to quickly approximate the effective resistances remains an open problem~\cite{spielman2011graph, koutis2012improved}.
Our choice is to adopt degree sampling. For an edge $e=(u, v)$, we set the sampling probability
$p_e \leftarrow \min(1, C A_{u,v} (d_u^{-1} + d_v^{-1}))$.
Here $d_u = \sum_{v} A_{u,v}$ is the degree of $u$. The following theorem from Lov\'asz et al.~\cite{lovasz1993random} showed
that the quantity $d_u^{-1} + d_v^{-1}$ is a simple but good upper bound to the effective resistance, especially for expander graphs:
\begin{theorem}[Corollary 3.3 in \cite{lovasz1993random}]
  \label{thm:ER}
  For  $\forall u, v\in V$,
  $
    \frac{1}{2} \left(\frac{1}{d_u}+\frac{1}{d_v}\right)\leq R_{u,v}  \leq \frac{1}{1-\lambda_2}\left(\frac{1}{d_u}+\frac{1}{d_v}\right)
  $, where $1-\lambda_2$ is the spectral gap of the normalized graph Laplacian.
\end{theorem}
\noindent Such a scheme ensures that the total number of edges
kept in expectation is $O(nC)$: we can show that for any
vertex $u$, we have $\sum_{v} A_{uv} d_u^{-1} = 1$
by the definition of $d_u$.
Furthermore, we can increase
the constant $C$
in order to increase the concentration of the samples we pick. In this work, we set $C=\log(n)$.
Experimentally, the above downsampling method has negligible effects on the
qualities of the embedding we produce, but significantly
reduces the edge count: most of our graphs initially have
at least $10$ times as many edges as vertices, and the random
walk graphs have even more edges.
We believe this can be justified using the guarantees of
this degree-sampling scheme in well-connected cases~\cite{SpielmanT11}.
For example, the spectral gap~(i.e., $1-\lambda_2$ in Theorem~\ref{thm:ER})  of the Blogcatalog graph~\cite{tang2009relational} is about $0.43$~\cite{qiu2020matrix},
and it is widely believed that most web graphs are well
connected, too~\cite{mohaisen2010measuring}.

Overall, the above step provides an $O(n\log{n})$-sparse but accurate estimator to the NetMF matrix in \Eqref{eq:NE}. 

\diff{
\subsection{Fast Randomized SVD}
\label{subsec:frsvd}

After constructing the sparsifier, the next step is to efficiently perform randomized SVD on the sparse matrix $\bm{M}$ and obtain the initial embedding. 
NetSMF and \lightne{} conduct randomized SVD algorithm on the sparsifer such that
$
  \mM \approx \mU \bm{\Sigma} \mV^\top
$, and define the embedding matrix  as  $\mX = \mU \bm{\Sigma}^{1/2}$.
\lightneplus{} improves the vanilla randomized SVD used in NetSMF and \lightne{} by incorporating power iteration scheme and fast orthonormalization.

Before formally introducing our techniques, we briefly review the core idea of randomized SVD~\cite{randnla,musco2015nips,HalkoMT11}.
 The method for randomized SVD mainly relies on the random projection to identify the subspace capturing the dominant information of input matrix $\bm{M}$, which can be realized by multiplying $\bm{M}$ with a random matrix on its right or left side to obtain the subspace’s orthonormal basis matrix $\bm{Q}$~\cite{HalkoMT11}. Then, the low-rank approximation in form of 
$\bm{Y}\bm{Q}^{\top}=\bm{M}\bm{Q}\bm{Q}^{\top}$ is computed with orthonormal basis matrix $\bm{Q}$. The approximate truncated SVD result of $\bm{M}$ can be generated by performing SVD on the reduced matrix $\bm{Y}$. Because $\bm{Q}$ has much fewer columns than $\bm{M}$, it reduces the computational time. The above theoretical basis constitutes the vanilla randomized SVD algorithm used in \lightne{}~\cite{qiu2021lightne}.

However, the vanilla randomized SVD algorithm in \lightne{} suffers from poor scalability and slow speed. Therefore, we review the related work~\cite{frpca} and find that several techniques are useful for improving the randomized SVD algorithm:

\vpara{Power Iteration Scheme} With power iteration scheme in~\cite{HalkoMT11}, the accuracy of the approximation can be further improved. The power iteration scheme is based on the fact that $(\bm{M}\bm{M}^{\top})^{q}\bm{M}$ has the same singular vectors as $\bm{M}$, but the singular values decay rapidly. Therefore, the orthonormal basis matrix $\bm{Q}$ produced from $(\bm{M}\bm{M}^{\top})^{q}\bm{M}$ makes a better approximation. Based on the fact that $\bm{M}$ is a sparse symmetric matrix generated by sparsifier, power iteration scheme can be simplified as $(\bm{M})^{q}\bm{M}$. Omitting the calculation of multiplying $\bm{M}^{\top}$ by a tall-and-thin dense matrix in power iteration can speed up the process when $\bm{M}$ is stored in row-major CSR format.

\vpara{Fast Orthonormalization Operation} The orthonormalization operation is used in power iteration after every sparse matrix-matrix multiplication to alleviate the round-off error in the floating-point computation. QR factorization will be the default choice for orthonormalization. LU factorization and eigSVD are two alternatives\cite{frpca}. 
In particular, eigSVD~(as summarized in Algorithm~\ref{alg:eigSVD}) is an economic SVD method, it can replace QR factorization because  left singular vectors are orthonormal.
In this work, we choose eigSVD for orthonormalization, mainly due to its efficiency.
Suppose we want to orthonormalize an $n\times d$ matrix, the time complexity of eigSVD is $O(C_{mul}nd^2+C_{eig}d^3)$, while QR factorization costs $O(C_{qr}nd^2)$ and LU factorization costs $O(C_{lu}(nd^2-d^3/2))$.
According to \cite{frpca}, the constant $C_{mul}$ is much smaller than $C_{qr}$ and $C_{lu}$, which makes eigSVD runs much faster than QR and LU, especially when $n\gg d$.
For example, to orthonormalize a tall-and-thin matrix with 1.72 billion rows and 32 columns, the QR factorization takes 1900 seconds, LU factorization takes 440 seconds, while eigSVD only takes 15 seconds. Overall, eigSVD will be our choice for orthonormalization in power iteration as it is the most computationally efficient compared to other methods. 
Besides orthonormalization in power iteration, we also replace all possible vanilla SVD with eigSVD in our algorithm.

\hide{In addition, eigSVD, as an economical SVD method, can replace the naive SVD method. At the last stage of power iteration, that is, Line 7-9 in Algo.~\ref{alg:frsvds}, we produce the approximate SVD of input symmetric matrix $\bm{M}$.}


\begin{algorithm}[t]
	\caption{eigSVD}
	\small
	\label{alg:eigSVD}
    \SetAlgoLined \DontPrintSemicolon
    \SetKwInOut{Input}{Input}
    \SetKwInOut{Output}{Output}
    \SetKwFunction{eigSVD}{\diff{eigSVD}}
    \SetKwProg{myalg}{Procedure}{}{}
    \diff{
    \myalg{\eigSVD{$\bm{X}$}}{
	  $\bm{C}=\bm{X}^{\top}\bm{X}$\tcp*{cblas\_sgemm}
	  $[\bm{V},\bm{D}] = \mathrm{eig}(\bm{C})$\tcp*{LAPACKE\_ssyevd}
	  $\bm{S}=\mathrm{sqrt}(\mathrm{diag}(\bm{D}))$\;
	  $\hat{\bm{S}}=\mathrm{spdiags}(1./\bm{S},0,n,n)$\;
	  $\bm{U}=\bm{X}\bm{V}\hat{\bm{S}}$\tcp*{cblas\_sgemm}
	  \KwRet $\bm{U},\bm{S},\bm{V}$\;}
  }{}
  \setcounter{AlgoLine}{0}
\end{algorithm}

Combining vanilla randomized SVD algorithm in \lightne{} with the above two techniques, we can propose a fast randomized SVD algorithm as Algorithm~\ref{alg:frsvds}.

\begin{algorithm}[t]
	\caption{Fast Randomized SVD}
	\small
	\label{alg:frsvds}
    \SetAlgoLined \DontPrintSemicolon
    \SetKwInOut{Input}{Input}
    \SetKwInOut{Output}{Output}
    \SetKwFunction{frsvds}{\diff{FastRandomizedSVD}}
    \SetKwProg{myalg}{Procedure}{}{}
    \diff{\myalg{\frsvds{$\bm{M}$, $d$, $q$}}{
        \tcc{We assume the input matrix $\bm{M}$ is symmetric.}
      $\bm{\Omega}=\mathrm{randn}(n,~d+s)$\tcp*{vsRngGaussian}
      $\bm{Y}=\bm{M}\bm{\Omega}$\tcp*{mkl\_sparse\_s\_mm}
	  $[\bm{Q},\_,\_] = \mathrm{eigSVD}(\bm{Y})$\;
	  \For{$i=1,2,...,q$}{ 
	      $\bm{Y}=\bm{M}\bm{Q}$\tcp*{mkl\_sparse\_s\_mm}
	      $\bm{T}=\bm{Q}$\;
		  $[\bm{Q},\bm{\Sigma},\bm{V}] = \mathrm{eigSVD}(\bm{Y})$\;
	  }
 	  $\bm{U}=\bm{Q}(:,1:d),\bm{\Sigma}=\bm{\Sigma}(1:d,1:d),\bm{V}=\bm{TV}(:,1:d)$\tcp*{cblas\_sgemm}
 	  \KwRet $\bm{U},\bm{\Sigma},\bm{V}$\;}
  }{}
  \setcounter{AlgoLine}{0}
\end{algorithm}

In Algo.~\ref{alg:frsvds}, $\bm{\Omega}$ is by default a Gaussian i.i.d matrix. Note that $\bm{\Omega}$ can also be a very sparse random projection~\cite{li2006very} or sparse sign random projection~\cite{tropp2019streaming} for better efficiency. 
Lines 5-8 is the power iteration scheme and eigSVD directly produce the approximate SVD of input symmetric matrix $\bm{M}$.
The hyperparameter $s$ in Line 2 is the oversampling parameter which enables $\bm{\Omega}$ with more than $d$ columns for better accuracy, which can be a smaller number such as $10$ or $20$ compared to embedding dimension $d=128$. 
The hyperparameter $q$ in Algo.~\ref{alg:frsvds} controls the number of power iterations. 
According to \cite{HalkoMT11}, the proposed fast randomized SVD algorithm has the following guarantee:

\beq{
  \label{eq:errorwithq}
  \besp{
  &\mathop{\mathbb{E}}\Vert\bm{M}-\bm{M}\bm{Q}\bm{Q}^{\top}\Vert_2 \\ 
  \leq&
  \left[1+\sqrt{\frac{d}{s-1}}+\frac{e(\sqrt{d+s})}{s}\sqrt{n-d}\right]^{\frac{1}{q+1}}\sigma_{d+1},
 }}where $\mathop{\mathbb{E}}$ denotes expectation, $e$ is natural constant and $\bm{Q}$ is the orthonormal matrix after the last step of power iteration. When we increase the power iteration parameter $q$, the power iteration scheme drives the extra factor in \Eqref{eq:errorwithq} to one exponentially fast.
}

\subsection{Spectral Propagation}
\label{subsec:spectral_prop}
Once the embedding $\mX$ is obtained, we
apply spectral propagation to further improve its quality. Following ProNE, the final embedding
is enhanced by applying a polynomial $\sum_{r = 0}^{k} c_{r} \mathcal{\mL}^{r} \mX$, 
where $k$ is the number of spectral propagation steps and by default set to be 10,
$\mathcal{\mL} \triangleq \mI - \mD^{-1}\mA$ is the normalized graph Laplacian matrix and $c_r$'s are the coefficients of Chebyshev polynomials. 



\diff{
\subsection{Automated Hyperparameter Tuning}
\label{sec:auto}
\hide{
The fast randomized SVD is more scalable and faster than the vanilla randomized SVD used in \lightne{}, which enables \lightneplus{} to embed a very large graph with 1B vertices and 100B edges in half an hour.
The low cost and high speed of \lightneplus{} allows us to add a further hyperparameter tuning step towards better embedding quality.
We propose to tune parameters for random walk matrix polynomials to obtain a better embedding. Then the \lightneplus{} make a spectral approximation of $\sum_{r=1}^T s_r(\bm{D}^{-1}\bm{A})^r \bm{D}^{-1}$, $s_r$ is the coefficients of random walk matrix polynomials.
}

Our design of \lightneplus{} includes several hyperparameters such as the number of edge samples in sparsifier construction (\Secref{subsec:algo_sparsifier}), the number of power iterations in fast randomized SVD (\Secref{subsec:frsvd}), and the propagation coefficients in spectral propagation (\Secref{subsec:spectral_prop}).
These hyperparameters play important role in the quality of network embedding. Thus there is a strong demand to automatically and quickly determine the best set of hyperparameters.
Fortunately, with the lightweightness and efficiency of \lightneplus{}, we are able to run \lightneplus{} multiple times with different hyperparameters to find the best hyperparameter configuration.

In this work, we employ a fast and and lightweight AutoML library FLAML~\cite{wang2021flaml} with BlendSearch~\cite{wang2021blendsearch} 
to automatically search hyperparameters of \lightneplus{},
and define the objective function of FLAML as the performance of learned network embedding on downstream task (such as F1 score in node classification tasks). 
\hide{
In particular, the BlendSearch strategy in FLAML combines local search with global search, which leverages the space exploration ability of global search methods such as Bayesian optimization~\cite{snoek2012bayesianopt}.  
Bayesian optimization is a common and powerful method to solve hyper-parameter optimization problems. The basic idea of Bayesian optimization is to use the Bayesian rule to estimate the posterior distribution of the objective function based on dataset, and then select the next sampling hyper-parameter combination according to the distribution. It aims to find the parameters that achieve the greatest improvement by optimizing the objective function or loss function. 
BlendSearch also requires a low-cost initial point as input if such point exists, and starts the search from there which speed up the search process. 
}
%
We include the following four categories of hyperparameters in the search space of FLAML:

\vpara{Random-walk Matrix Polynomial Coefficients $s_r$'s}
Previous work has noticed that the coefficients of matrix polynomial play an important role in network embedding~\cite{abu2018watch,chen2019fast}. Thus we 
search the coefficients of random-walk matrix polynomials, i.e., $s_1, \cdots s_T$ in \Eqref{eq:NE}. The search space of each $s_r$ is set to be uniform from $[0.01, 1]$, and then the coefficients are normalized to satisfy $\sum_{r=1}^T s_r = 1.$

\vpara{The Number of Edge Samples $M$} The literature of random-walk matrix polynomial sparsification~\cite{cheng15sample,cheng2015spectral} points out that sampling more edge samples reduces the approximation error of sparsification. Although we observed this phenomenon in most real-world datasets, our experiments in Hyperlink2014-Sym graph (as discussed later  in \Secref{subsec:verylarge}) show that excessive edge samples can sometimes bring negative effects in downstream tasks.
Thus it is more reasonable to automatically search the proper 
number of edge samples to achieve better performance in downstream tasks. We allow FLAML to explore the number of edge sample in a log-uniform way.

\vpara{The Number of Power Iterations $q$} 
\Eqref{eq:errorwithq} suggests that increasing the number of power iterations can reduce the expected approximation error of fast randomized SVD.
However, we do find some cases (e.g., the case of Clueweb-Sym as discussed later in \Secref{subsec:verylarge}) where too many power iterations slightly hurt the network embedding quality in downstream tasks. Thus, we allow FLAML to search the best power iteration parameter $q$.

\vpara{Spectral Propagation Hyperparameters}
Inspired by \cite{hou2021automated}, we also search the hyperparameters of spectral propagation, especially the propagation steps $k$, and also parameters $\theta$ and $\mu$ which control the coefficients $c_r$'s. We refer interested readers to the ProNE paper~\cite{zhang2019prone} and its codebase\footnote{\url{https://github.com/THUDM/ProNE}} for a more detailed discussion on these hyperparameters.
}

%% file: 4.system.tex
\section{L\texorpdfstring{\MakeLowercase{\textsc{ight}}}{ight}NE 2.0: System Design}
\label{sec:sys}

\vpara{Overview}
We make a series of system optimizations to enable \lightneplus{} in a CPU-only shared-memory machine.
As introduced in \Secref{subsec:algorithm}, \lightneplus{} consists of two steps --- NetSMF and spectral propagation.
From a system perspective, the NetSMF step can be further decomposed into two sub-steps ---
parallel sparsifier construction
and \diff{parallel fast randomized SVD.}
In this section, we present our acceleration techniques for these components.
In \Secref{subsec:gps}, we introduce a new graph processing system, GBBS, which we leverage throughout \lightneplus{}. It includes the graph primitives and compression techniques.
In \Secref{subsec:buildsparsifier}, we discuss how we optimize
sparsifer construction with  GBBS and sparse parallel hashing,
which enable us to aggregate and construct
the sparsifier in memory efficiently.
\diff{Lastly, in \Secref{subsec:mkl}, we describe our fast randomized SVD and spectral propagation implementation using Intel MKL.}
An overview of our design
can be found in \Figref{fig:system_overview}.
\subsection{Sparse Parallel Graph Processing}\label{subsec:gps}
\hide{
  Both the sparsification of random walk molynomials described in \Secref{subsec:NetSMF}
  and the initial embedding of ProNE in \Secref{subsec:PRONE} require building a sparse
  matrix from a graph.
  Furthermore, we note that most real-world graphs are highly sparse.
  Indeed, $m/n$, or the average degree of nearly all of the networks available from the SNAP~\cite{leskovec2014snap} and LAW~\cite{BoVWFI} datasets is below 100, with the average ratio of $m/n$ being close to 20.
}
\lightneplus{} involves intensive graph operations, such as
performing random walks, querying vertex degrees, random accessing a neighbor
of a vertex, etc. In this work, we build on the Graph Based Benchmark
Suite~(GBBS)~\cite{dhulipala2018theoretically}, which extends the Ligra~\cite{shun2013ligra} interface with additional
purely-functional primitives such as maps, reduces, filters over both vertices and
graphs. We chose GBBS because it is performant, relatively simple to use, and has already been shown to scale to real-world networks with billions--hundreds of billions of edges on a single machine, achieving state-of-the-art running times for many fundamental graph problems.

\vpara{Compression}
An important design consideration for \lightneplus{} is to embed very large graphs on a single machine. Although the CSR format is normally regarded as a good compressed graph representation~\cite{kepner2011graph}, we need to further compress this data structure and reduce memory usage.
Our approach builds on  state-of-the-art parallel graph compression techniques, which enable both fast parallel graph encoding and decoding.
In particular, we adopt the \textit{parallel-byte} format from Ligra+~\cite{shun2015ligraplus}. 
In \textit{sequential byte coding}, we store a vertex's neighbor list by difference encoding consecutive vertices, with the first vertex difference encoded with respect to the source.
A decoder processes each difference one at a time, and sums the differences into a running sum which gives the ID of the next neighbor.
Unfortunately, this process is entirely sequential, which could be costly for high-degree vertices and thus inhibit parallelism.
The \textit{parallel-byte} format from Ligra+ breaks the neighbors of a high-degree vertex into blocks, where each block contains a configurable number of neighbors.
Each block is internally difference-encoded with respect to the source.
As each block can have a different compressed size, the format also stores offsets from the start of the vertex to the start of each block. In what follows, when we refer to \emph{compressed} graphs, we mean graphs in the CSR format where neighbor lists are compressed using the parallel-byte format.

To the best of our knowledge, we are the first to introduce GBBS and Ligra+ to the network embedding problem.
\hide{
  An attractive feature of this approach is that the \lightne{}'s code is high-level, enabling simplicity and understandability, since all graph manipulations are implemented through a clean, high-level API.
  Furthermore, this interface completely hides the details of the underlying graph's storage from the user, which enables the same code to apply without any modifications to both uncompressed and parallel-byte compressed CSR networks.
}



\subsection{Parallel Sparsifier Construction}
\label{subsec:buildsparsifier}

As described in Section~\ref{sec:ne}, building the sparsifier requires:
(1) generating a large number of edge samples using the PathSampling in Algo.~\ref{alg:pathsample}
and (2) aggregating the sampled edges to count the frequency each distinct edge appears.
After ensuring that the input graph is compressed, the main challenge in our design is to efficiently construct and
store the sparsifier in memory.
In this section, we use both the purely-functional primitives and the parallel compression techniques in GBBS to  scalably and memory-efficiently conduct the PathSampling.
We further employ sparse parallel hashing to aggregate the sampled edges and construct the sparsifer.



\vpara{Parallel Per Edge PathSampling by GBBS}
A natural idea is to repetitively 
call PathSampling$(G, u, v, r)$ (Algo.~\ref{alg:pathsample})
with a uniformly sampled edge
$(u, v)$ and a uniformly sampled path length $r\in [T]$.
Unfortunately, this approach is challenging to implement on compressed graphs 
--- it requires an efficient way to sample and access a random edge. 
Straightforward methods store all edges in an array which enables $O(1)$ random access, or perform binary search on the prefix sums of vertex degrees and select the chosen edge incident to a particular vertex.
The former would require a prohibitive amount of memory for our largest networks, and the latter would require extra $O(\log{n})$ time for binary searching each sample. 
\begin{algorithm}[t]
  \caption{Downsampled Per-Edge PathSampling.}
  \label{alg:geomsample}
  \footnotesize
  \SetAlgoLined \DontPrintSemicolon
  \SetKwInOut{Input}{Input}
  \SetKwInOut{Output}{Output}
  \SetKwFunction{peredgesample}{DownSampledPerEdgePathSampling}
  \SetKwBlock{MapEdges}{$G$.\textsc{MapEdges}(\textsc{function} $(e = (u,v)) \rightarrow$}{)}{}
  \SetKwBlock{Lambda}{}{}
  \SetKwProg{myalg}{Procedure}{}{}
  \SetKwProg{myfunc}{Function}{}{End Function}
  \myalg{\peredgesample{$G$, $T$}}{
  \MapEdges()
  {
  $n_{e} \leftarrow  \lfloor M/m \rfloor + \text{random variable from Bernoulli}\left(\{M/m\}\right)$\;
  \For{$i \gets 1$ to $n_e$}
  {
  $p \leftarrow \text{Uniform}[0, 1]$\; 
  \diff{$r \leftarrow \text{Multinomial}[s_1, \cdots, s_T]$\;} 
  \If{$p < p_e$}
  {
  $(u', v') \leftarrow$ PathSampling($G, u, v, r$)\;
  Add $(u', v')$ with weight $1/{p_e}$ to the sparsifier\textbf{)}\;
  }
  }
  }
  }{}
\end{algorithm}

Instead, we propose Algo.~\ref{alg:geomsample}, which describes an equivalent process that has the benefit of being more cache and memory-friendly, and works seamlessly alongside compression.
The idea is to map over the edges in parallel, and for each edge $e=(u,v)$ we run PathSampling~(Algo.~\ref{alg:pathsample}) $n_e$ times where $n_e$ is $ \lfloor M/m \rfloor$ plus a Bernoulli random variable with mean $\{M/m\}.$\footnote{$\lfloor \cdot \rfloor$ is the floor function, and $\{\cdot\}$ is the fractional part of a number.}
Since each edge $e$ is sampled independently, the expected number of samples is exactly $M$.\footnote{$\E[n_e]=M/m$, and $\E[\sum_{e\in E}n_e] = \sum_{e\in E}\E[n_e] = M$.}
And it is easy to see that $\Theta(M)$ samples are drawn with high probability by standard concentration bounds.
After sampling a value $n_{e}$ from this random variable for a given edge $(u,v)$, we perform $n_{e}$ many random walks from $(u,v)$ treating it as though it had been selected uniformly at random in the original process.
We further incorporate the edge downsampling~(as introduced in \Secref{subsec:algorithm}) into Algo.~\ref{alg:geomsample}  to reduce the sample complexity. After drawing random variable $n_e$~(Algo.~\ref{alg:geomsample}, Line 3), for each of the $n_e$ times this edge $e$ is sampled, we flip a coin that comes up heads with  probability $p_e$~(Algo.~\ref{alg:geomsample}, Line 5). We then apply Algo.~\ref{alg:pathsample} and add the sampled edge pair to the sparsifier with adjusted weight $1/p_e$ only if the coin comes up heads~(Algo.~\ref{alg:geomsample}, Line 8-9). \diff{The parameter $r$ in Line 6 is sampled from a mutinomial distribution whose parameters are the  coefficients of random walk matrix polynomial (i.e., $s_1, \cdots, s_T$).}
Our implementation of the above idea uses the \textsc{MapEdges} primitive~(Algo.~\ref{alg:geomsample}, Line 2) in GBBS, which applies a user-defined function over every edge in parallel.
The user-defined function~(Algo.~\ref{alg:geomsample}, Line 3-9)  conducts the downsampled per-edge sampling  we introduced above.

Lastly, we observe that implementing random walks in the shared-memory setting requires efficiently fetching the $i$-th edge incident to a vertex during the walk.
This is because we simulate the random walk one step at a time by first sampling a uniformly random 32-bit value, and computing this value modulo the vertex degree.
Fetching an arbitrary incident edge is trivial to implement for networks stored in CSR~(without extra compression) by simply fetching the offset for a vertex and accessing its $i$-th edge.
However, for graphs in CSR where adjacency information is additionally compressed in the parallel-byte format, we may need to decode an \emph{entire block} in order to fetch the $i$-th edge.
To help mitigate this cost, we chose a block size of $64$ after experimentally evaluating the trade-off between the compressed size of the graph in memory, and the latency of fetching arbitrary edges incident to vertices.
We note that further optimizations of this approach, such as batching multiple random walks accessing the same (or nearby vertices) together, to mitigate the cost of accessing these vertices' edges would require a careful analysis of the overhead for shuffling the data via a semisort~\cite{semisort}, or a partial radix-sort~\cite{milk} vs.~the overhead for performing random reads. Optimizations of this flavor to further improve locality may be an interesting direction for future work.


\vpara{Sparse Parallel Hashing}
Next, we turn to how the sparsifier is constructed and represented in memory.
After running Algo.~\ref{alg:geomsample} which generates many weighted edges, we need to count the frequency each distinct edge is sampled.
We considered several different techniques for this aggregation problem in the shared-memory setting, including (1) generating per-processor lists of the edges and then merging the lists using the efficient sparse-histogram  introduced in GBBS~\cite{dhulipala2018theoretically} and (2) storing the edges and partial-counts in per-processor hash tables that are periodically merged.

Ultimately, we found that the fastest and most memory-efficient method across all of our inputs was to use sparse parallel hashing.
The construction used in this paper is folklore in the parallel algorithms literature, and we refer to Maier et al.~\cite{maier2016concurrent} for a detailed explanation of the folklore algorithm.
In a nutshell, our parallel hash table stores a distinct entry for each edge that is ever sampled, along with a count.
Threads can access the table in parallel, and collisions are resolved using linear probing.
Note that we do not require deletions in this setting.
When multiple samples are drawn for a single edge, the counts are atomically incremented using the atomic \textsc{xadd} instruction.
We note that the \textsc{xadd} instruction is significantly faster than a more naive implementation of a \textsc{fetch-and-add} instruction using \textsc{compare-and-swap} in a while loop when there is contention on a single memory location, and \textsc{xadd} is only negligibly slower in the light-load case~\cite{shun13reducing}.
Our implementation is lock-free and ensures that the exact count of each edge is computed, since our implementation uses atomic instructions to ensure that each sample is accounted for.

\vpara{Memory  Tricks}
Almost all data structures in \lightneplus{}, especially in the sparsifier construction step, require allocating large and contiguous chunks of memory. 
Thus we try our best to avoid the memory fragmentation problem so that
we can maximize memory usage.
For example,
we conduct element-wise logarithm in-place in the hash table
and convert the hash table~(in COO format)
to the CSR sparse matrix for fast randomized SVD.

\subsection{High Performance Linear Algebra}
\label{subsec:mkl}

We use Intel MKL to optimize linear algebra operations in \lightneplus{}, which appear in fast randomized SVD and spectral propagation frequently.

\diff{
\vpara{Fast Randomized SVD} After constructing the sparsifier, the next step is to efficiently perform fast
randomized SVD and obtain the initial embedding.
The fast randomized SVD, as described in Algo.~\ref{alg:frsvds} involves excessive linear algebra operations,
which are well-supported and highly-optimized by the Intel MKL library.
For example, its random projection
is, in essence, a product of an $n\times n$ sparse matrix
and a dense $n\times d$ Gaussian
random matrix, which are implemented in  MKL's Sparse BLAS Routines.
Other examples include conducting eigSVD on the projected matrices (as described in Algo.~\ref{alg:eigSVD}), which are all supported by Intel MKL BLAS and LAPACK routines.
We list the pseudo-code of fast randomized SVD and eigSVD, as well as the corresponding Intel MKL routines in Algo.~\ref{alg:frsvds} and Algo.~\ref{alg:eigSVD} respectively.
}

\vpara{Spectral Propagation} Besides randomized SVD,
the spectral propagation step also involves
linear algebra operations.
Note that the spectral propagation step is highly efficient.
It does not need to evaluate the higher powers of $\mathcal{\mL}$,
but rather only applies repeated
Sparse Matrix-Matrix multiplication~(SPMM) between a sparse  $n\times n$
Laplacian matrix $\mathcal{L}$
and a dense $n \times d$ embedding matrix,
which can also be
handled by MKL Sparse BLAS routines.

\hide{
\subsection{Spectral Propagation}
\vpara{Spectral Propagation} Besides fast randomized SVD,
the spectral propagation step also involves
linear algebra operations.
Note that the spectral propagation step is highly efficient.
It does not need to evaluate the higher powers of $\mathcal{\mL}$,
but rather only applies repeated
Sparse Matrix-Matrix multiplication~(SPMM) between a sparse  $n\times n$
Laplacian matrix $\mathcal{L}$
and a dense $n \times d$ embedding matrix,
which can also be
handled by MKL Sparse BLAS routines.
\hide{
\vpara{Memory  Tricks}
Almost all data structures in \lightne{} require allocating large and contiguous chunks of memory. 
Thus we try our best to avoid the memory fragmentation problem so that
we can maximize memory usage.
For example,
we conduct element-wise logarithm in-place in the hash table
and convert the hash table~(in COO format)
to the CSR sparse matrix for MKL directly.}
}

%% file: 5.eval.tex
\section{End-to-end Evaluations}
\label{sec:eval}

\hide{ 
    \jz{comments from Chi:
        Basically, we need to be different from an ML paper
        We can partition the datasets into three groups by scale:
        the largest group, on which our system is the only one working;
        the 2nd largest group,
        which contain the largest datasets studied in previous work,
        and only a small number of systems (GraphVite etc.) are applicable;
        and the 3rd group, on which many methods including GNN are applicable
        I think our submission should mainly focus on the first two groups,
        and use the third group as a sanity check.
        But be clear that the third group is not our main target scenario
        for papers targeting the third group, they should be submitted to an ML conference
    }
}
\revise{
In this section,
we evaluate \lightneplus{} on
nine graph datasets, summarized in Table~\ref{tbl:data}.
These datasets fall into three natural groups by scale: (1) small-sized benchmarks standard to the network embedding literature, such as Blogcatalog and Youtube; (2) large graphs in previous works, such as Friendster studied by \gvt{} and OAG studied by NetSMF; (3) some of the largest publicly-available graphs where no previous results on network embeddings exist.
The small graphs from (1) are used to verify the effectiveness of \lightneplus{}, although they are not our main target scenarios. We compare \lightneplus{} with all the baselines in small graphs.
We use the large graphs from (2) to demonstrate the effectiveness and efficiency of our method. For large graphs, we compare with baselines on their corresponding example datasets in the original papers. The experiments on the very large graphs from (3) further demonstrate that our system can scale beyond previous work. As for the very large graphs, we found that all other baselines failed to learn embedding due to huge memory cost, or requiring a large number of GPUs or time.}

\hide{
    These datasets are partitioned into three groups by scale ---
    large graphs, very large graphs, and small graphs.
    The first group contains the largest datasets studied in previous work, such as Friendster studied by \gvt{} and OAG studied by NetSMF.
    We conduct thorough experiments on this group to illustrate the efficiency and scalability of our method.
    In the second group, we introduce some largest public-available graphs on which our system is the only one working.
    The third group includes small-scale and commonly adopted graphs such as BlogCatalog and YouTube,
    which we use as a sanity check for effectiveness  of \lightne{} and is not our main target scenario.
}
We set up our evaluation in \Secref{subsec:setup} and then report experimental results in the three groups, respectively. In this section, we refer \lightne{} to the system in~\cite{qiu2021lightne} and refer \lightneplus{} to the system in this work (i.e., the system enhanced by fast randomized SVD algorithm).  

\vspace{-0.2cm}
\subsection{Experimental Setup}
\label{subsec:setup}
\vpara{Hardware Configuration}
For \lightneplus{}, all experiments are conducted on a server
with two Intel\textsuperscript \textregistered Xeon\textsuperscript \textregistered E5-2699~v4 CPUs~(88 virtual cores in total) and 1.5~TB memory.
\begin{table}[tb]
    \caption{
        Hardware configurations
        and their most similar counterparts in Azure.
        N/A indicates it is not reported in the original paper.
    }
    \label{tbl:azure}
    \centering
    \small
    \begin{tabular}{l@{~}|l@{~}|r @{~}r  @{~}r @{~}r}
        \toprule
                                &             & vCores & RAM       & GPU     & Price~(\$/h) \\\midrule
        \multirow{5}{*}{System} & \gvt{}      & N/A    & 256~GB    & 4X~P100 & N/A          \\
                                & \shortpbg{} & 48     & 256~GB    & 0       & N/A          \\
                                & NetSMF      & 64     & 1.7~TB    & 0       & N/A          \\
                                & \lightne{}     & 88     & 1.5~TB    & 0       & N/A          \\
                                & \lightneplus{}  & 88     & 1.5~TB    & 0       & N/A         \\\midrule
        \multirow{4}{*}{Azure}  & NC24s~v2    & 24     & 448~GiB   & 4X~P100 & 8.28         \\
                                & E48~v3      & 48     & 384~GiB   & 0       & 3.024        \\
                                & M64         & 64     & 1024~GiB  & 0       & 6.669        \\
                                & M128s       & 128    & 2,048~GiB & 0       & 13.338       \\
        \bottomrule
    \end{tabular}
    \normalsize
\end{table}
\begin{table*}[htbp]
    \caption{Datasets statistics. \revise{Size represents the size of corresponding hard-disk file. * indictes that very large graphs are stored in compressed CSR format based on GBBS, while other datasets are stored in CSR format.}}
    \label{tbl:data}
    \centering
    \footnotesize
    \begin{tabular}{@{}r@{~}|@{~}r@{~}r@{~}|@{~}r @{~}r @{~}r  @{~}r @{~}r| @{~}r @{~}r@{}}
        \toprule
              & \multicolumn{2}{c|}{Small Graphs~($|E|\leq 10M$)} & \multicolumn{5}{c|}{Large Graphs~($10M<|E|\leq 10B$)} & \multicolumn{2}{c}{Very Large Graphs~($|E|>10B$)}                                                                                                       \\ \midrule
              & BlogCatalog                                       & YouTube                                               & LiveJournal                                       & Friendster-small & Hyperlink-PLD & Friendster    & OAG         & ClueWeb-Sym    & Hyperlink2014-Sym \\\midrule
        $|V|$ & 10,312                                            & 1,138,499                                             & 4,847,571                                         & 7,944,949        & 39,497,204    & 65,608,376    & 67,768,244  & 978,408,098    & 1,724,573,718     \\
        $|E|$ & 333,983     & 2,990,443      & 68,993,773                                        & 447,219,610      & 623,056,313   & 1,806,067,142 & 895,368,962 & 74,744,358,622 & 124,141,874,032   \\
        \revise{size} & \revise{3.2~MB}  & \revise{46~MB} & \revise{589~MB}    & \revise{6.7~GB}  & \revise{9.9~GB}   & \revise{31~GB} & \revise{16~GB} & \revise{107~GB*} & \revise{197~GB*} \\
        \bottomrule
    \end{tabular}
\end{table*}


\vpara{Accuracy Metrics}
We follow the tasks and evaluation metrics in the original proposals.
When comparing to \shortpbg{} on LiveJournal, we evaluate the link prediction task with metrics to be mean rank~(MR), mean reciprocal rank~(MRR), and HITS@10.
When comparing to \gvt{} on Hyperlink-PLD, we evaluate the
link prediction task with metric to be  AUC. For the rest of the datasets,
the task is node classification. After generating the embedding, we randomly sample a portion of labeled vertices for training and use the remaining for testing. The task is completed by the one-vs-rest logistic regression model, which is implemented by LIBLINEAR~\cite{fan2008liblinear}. We repeat the prediction procedure ten times and evaluate the average performance in terms of both Micro-F1 and Macro-F1 scores~\cite{tsoumakas2009mining}. 

\vpara{Efficiency Metrics}
We compare both the time and cost efficiency of different systems.
Time efficiency is measured by running time, 
while cost efficiency is measured by estimated cost. 
Our cost estimation is based on the pricing~(\$) on Azure Cloud~(The AWS price is very similar).
We search the most suitable Azure instance for each system and then use its price per hour multiplied by the running time to estimate the cost.
As shown in Table~\ref{tbl:azure}, we assume
\gvt{} uses NC24s~v2, \shortpbg{} uses E48~v3,
while NetSMF, \lightne{}, and \lightneplus{} use M128s.
The reason why we present cost efficiency is
that different systems have different hardware requirements
--- \gvt{} is a CPU-GPU hybrid system, while
\lightne{}, \lightneplus{} and NetSMF are CPU applications.
Thus we use cloud rent price as a measure for the value of different hardware configurations.

\subsection{Small Graphs}
\label{subsec:small}
\begin{figure}[t]
    \centering
    \includegraphics[width=0.48\textwidth]{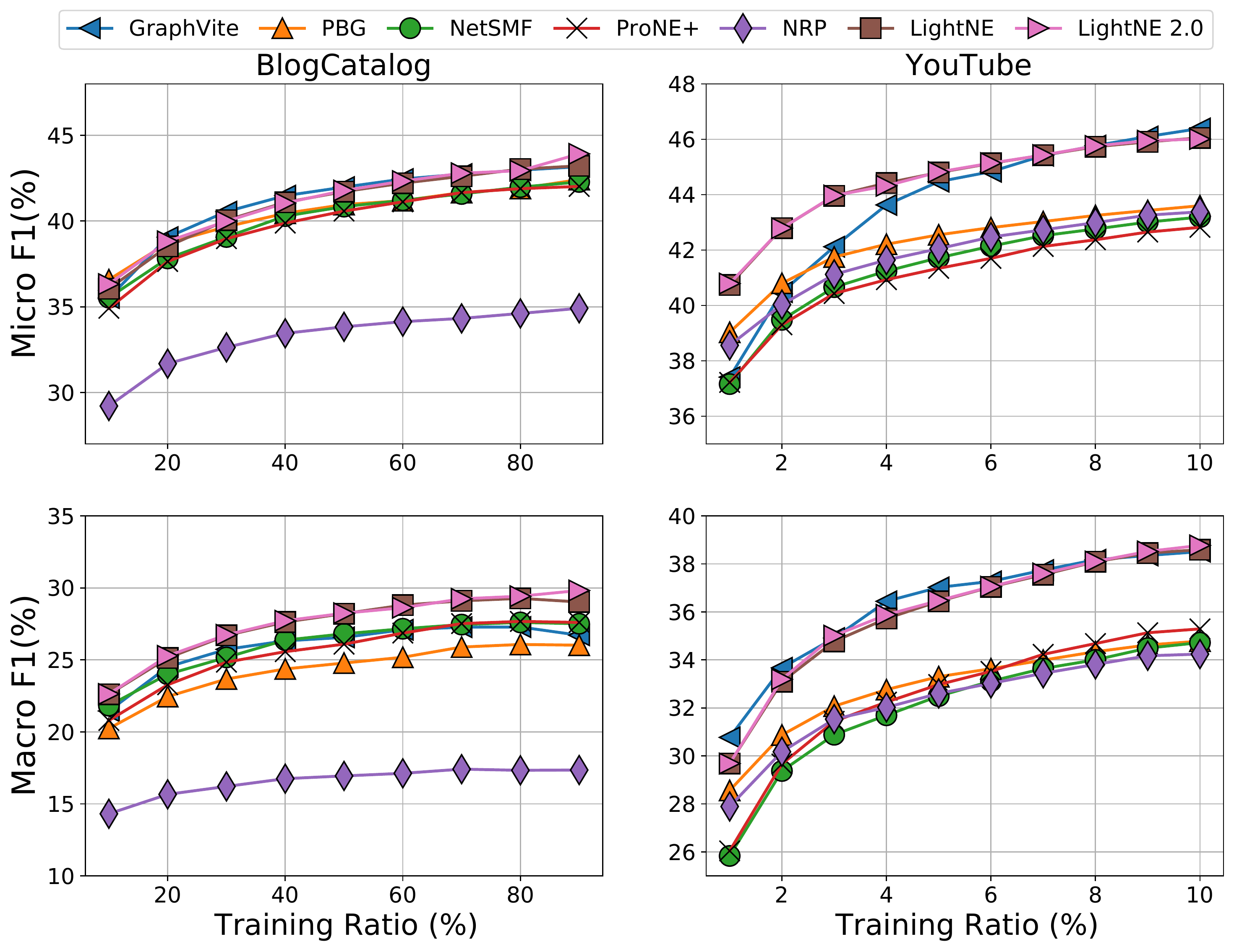}
    \caption{\revise{Node classification results on small graphs.}}
    \label{fig:small_graph}
\end{figure}

As shown in \Figref{fig:small_graph},
we compare the prediction performance of \lightneplus{} against
all the baselines\footnote{NRP is from \url{github.com/AnryYang/nrp}.
\shortpbg{}'s hyper-parameters for YouTube haven't been officially released, so we set them by cross-validation.} in BlogCatalog and YouTube. The Python implementation released by ProNE paper~\cite{zhang2019prone} is inefficient, and it ``requires 29 hours to embed a network of hundreds of millions of nodes''~(quote from the abstract of ProNE paper~\cite{zhang2019prone}). For fair comparison, we use the re-implemented ProNE in \lightne{} which is enhanced by system optimizations~(highly optimized GBBS for graph processing and MKL for linear algebra operations). The re-implementation has comparable accuracy to the original one on datasets used in ProNE~\cite{zhang2019prone}, but much faster. We refer to this new implementation as ProNE+.
In BlogCatalog, \lightneplus{} outperforms all the baselines consistently in terms of Macro-F1, and achieves comparable results to  \gvt{} regarding Micro-F1.
In YouTube,  the right panel of \Figref{fig:small_graph} suggests
that \lightneplus{}, together with \gvt{}, consistently yields  the best results among all the
methods. In particular, \lightneplus{} shows better Micro-F1 than \gvt{} when the training ratio is small~(1-6\%).
We also highlight that ProNE+ performs consistently worse than \lightneplus{}, showing that enhancing a simple embedding via spectral propagation may yield sub-optimal performance. \diff{\lightneplus{} is also slightly better than \lightne{} (on average 0.49\% and 0.24\% relatively better on BlogCatalog and YouTube in terms of Macro F1).}

The experiments on small graphs demonstrate the effectiveness of our system, though the system is mainly designed for larger graphs.

\subsection{Large Graphs}
\label{subsec:large}
We compare to four systems that are designed for large graphs: \pbg, \gvt, NetSMF and \lightne{}. 
We use the tasks, datasets, hyper-parameters, and evaluation
scripts provided by the corresponding papers’ GitHub repos in making these comparisons.

\subsubsection{Comparison with \pbg~(\shortpbg{})}
\label{subsec:pbg}

We compare with \shortpbg{} on the LiveJournal dataset~\footnote{
    \shortpbg{}
    reports results on LiveJournal, YouTube and Twitter, but
    only
    releases configuration for LiveJournal
    in the official github repository.
}.
For embedding dimension, all methods use 1024, which is the preferred value from \shortpbg{}. 
By cross-validation, we set $T=5$ for \lightne{}, and $T=5,q=1$ for \lightneplus{}.
%
%
The results are reported 
in Table~\ref{table:LiveJournal}.

\begin{table}[t]
    \caption{\revise{Comparison results of link prediction for \lightneplus{} and \shortpbg{} on LiveJournal.}}
    \label{table:LiveJournal}
    \small
    \centering
    \begin{tabular}{@{~}l|rrrrr@{~}}
        \toprule
                  & Time            & Cost            & MR            & MRR           & Hits@10       \\\midrule
        \shortpbg & 7.25~h          & \$21.95         & 4.247          & 0.874         & 0.929          \\
        \lightne  & 16.0~min & \$3.56 & 2.130 & \textbf{0.912} & \textbf{0.977} \\
        \lightneplus{} & \textbf{14.4~min} & \textbf{\$3.20} & \textbf{2.115} & \textbf{0.912} & \textbf{0.977} \\
        \bottomrule
    \end{tabular}.
    \normalsize
\end{table}
Not only does \lightneplus{} achieve better performance regarding all metrics,
but it also reduces time and cost by one order of magnitude. \diff{Specifically,
\lightneplus{} is 30$\times$ faster
and 7$\times$ cheaper than \shortpbg{}. Moreover, \lightneplus{} is faster than \lightne{} (14.4 min v.s. 16 min) due to the proposed fast randomized SVD algorithm.}

\subsubsection{Comparison with \gvt}

\gvt{} 
offers the evaluation of link prediction task on Hyperlink-PLD and node classification task on Friendster-small and Friendster.    
For embedding dimension, all methods use 128 on Hyperlink-PLD and Friendster-small, and use 96 on Friendster, which is the preferred value from \gvt{}.
\diff{By-cross-validation, we set $T=5$ for \lightne{} and \lightneplus{}, and additionally set $q=1$ for \lightneplus{}.
\lightneplus{} achieves AUC score 96.8 and outperforms \gvt{}'s 94.3.}
Friendster-small and Friendster are two datasets used for node classification tasks.
The performance on these two datasets is measured by Micro and
Macro F1 with varying ratio of labeled data. 
For \gvt{}, we use its preferred hyper-parameters.
For \lightne{} and \lightneplus{}, cross-validation shows that the best performance is obtained by setting $T=1$ and we additionally set $q=1$ for \lightneplus{}.
We list the performance when label ratio is 1\%, 5\%, and 10\% in Table~\ref{tbl:friendster}.
\begin{table}[t]
    \caption{\revise{Comparison results of node classification and link prediction between \lightneplus{} and \gvt{} on Friendster-small, Friendster and Hyperlink-PLD.
    FS, F and H stand for Friendster-small, Friendster and Hyperlink-PLD, respectively.
    }}
    \label{tbl:friendster}
    \centering
    \footnotesize
    \begin{tabular}{l|l|rccc}
        \toprule
        Metric                    & Dataset                           & Label Ratio~(\%) & 1              & 5              & 10             \\\midrule
        \multirow{6}{*}{Micro-F1} & \multirow{3}{*}{FS} & \gvt             & 76.93          & 87.94          & 89.18          \\
                                  &                                   & \lightne{}       & 84.53 & 93.20 & 94.04 \\
                                  &                                   & \lightneplus{}    & \textbf{86.25} & \textbf{93.53} & \textbf{94.24} \\
        \cmidrule{2-6}
                                  & \multirow{3}{*}{F}       & \gvt             & 72.47          & 86.30          & 88.37          \\
                                  &                                   & \lightne{}       & 80.72 & 91.11 & 92.34 \\
                                  &                                   & \lightneplus{}       & \textbf{84.11} & \textbf{91.92} & \textbf{92.72} \\
        \midrule
        \multirow{6}{*}{Macro-F1} & \multirow{3}{*}{FS} & \gvt             & 71.54          & 86.77          & 88.42          \\
                                  &                                   & \lightne{}       & 81.42 & 93.52 & 94.40 \\
                                  &                                   & \lightneplus{}      & \textbf{83.35} & \textbf{93.77} & \textbf{94.52} \\
        \cmidrule{2-6}
                                  & \multirow{3}{*}{F}       & \gvt             & 66.22          & 86.28          & 88.79          \\
                                  &                                   & \lightne{}       & 76.20 & 91.10 & 92.55 \\
                                  &                                   & \lightneplus{}{}       & \textbf{81.23} & \textbf{92.22} & \textbf{93.07} \\
        \midrule\midrule
        \multirow{3}{*}{AUC} & \multirow{3}{*}{H}  & \gvt{}           & 94.3           & & \\
            &         & \lightne       & 96.7 & & \\
             &       & \lightneplus{} &\textbf{96.8} &  & \\ 
        
        \bottomrule
    \end{tabular}.
\end{table}
\diff{
As we can see, \lightneplus{} is significantly better than \gvt{}. \lightneplus{} significantly outperforms \lightne{} due to applying the power iteration scheme in fast randomized SVD. It greatly improves the \lightne{}'s performance, especially when label ratio is equals to 1\% (Friendster-small and Friendster's Micro-F1 increased by 1.7 and 3.39, respectively).}


As for efficiency, 
the detailed  comparison among \lightne{}, \lightneplus{}, and \gvt{} is summarized in Table~\ref{tbl:gvt}.
\diff{
As we can see,
\lightneplus{} can embed 
Hyperlink-PLD in 19.68~min, 16$\times$ faster than \gvt{}.
Moreover, comparing to \gvt{},
\lightneplus{} achieves 31$\times$ and 84$\times$ speedup respectively in Friendster-small and Friendster,
and saves the cost by orders of magnitude~(24$\times$ cheaper on Friendster-small and 65$\times$ cheaper on Friendster).
Furthermore, the fast randomized SVD makes \lightneplus{} much more efficient than \lightne{}. For example, \lightneplus{} only takes 14.47 min to embed Friendster, 2.6x faster than \lightne{} which takes 37.6 min.
}

\begin{table}[t]
    \caption{Efficiency comparison between \lightneplus{} and \gvt{}. FS, F and H stand for Friendster-small, Friendster and Hyperlink-PLD, respectively.}
    \label{tbl:gvt}
    \centering
    \small
    \begin{tabular}{l|l|r|r|r}
        \toprule
                              &            & FS  & H      & F        \\\midrule
        \multirow{3}{*}{Time} & \gvt       & 2.79~h            & 5.36~h             & 20.3~h            \\
                              & \lightne{} & \textbf{5.83~min} & \textbf{29.77~min} & \textbf{37.60~min} \\
        & \lightneplus{} & \textbf{5.42~min} & \textbf{19.68~min} & \textbf{14.47~min}    \\
                              \midrule
        \multirow{3}{*}{Cost} & \gvt       & \$28.84           & \$44.38            & \$209.84          \\
                              & \lightne{} & \textbf{\$1.30}   & \textbf{\$6.62}    & \textbf{\$8.36}   \\
                              & \lightneplus{}& \textbf{\$1.20}   & \textbf{\$4.38}    & \textbf{\$3.22}   \\
        \bottomrule
    \end{tabular}.
\end{table}

\subsubsection{Comparing with NetSMF and ProNE+}
\label{subsec:cmp_netsmf_prone}
NetSMF and ProNE are redesigned and used as building blocks of both \lightne{} and \lightneplus{}.
In this section, we focus on
investigating how \lightneplus{} combines and strengthens the advantages of both. The dataset we adopt is OAG~\cite{sinha2015overview}, the largest graph from NetSMF.
We vary the number of edge samples $M$ conducted by \lightne{} and \lightneplus{} from $0.077Tm$ to $17Tm$.
We denote the configuration with the small number of edge samples ($M=0.077Tm$) as \lightne{}~(S) and \lightneplus{}~(S), and the one with the larger number edge samples~($M=17Tm$) as \lightne{}~(L) and \lightneplus{}~(L).
We find that power iteration parameter $q$ has a great impact on performance for \lightneplus{}. By balancing the runtime and performance, we report the result when choosing $q=5$ in Table~\ref{tbl:oag}.
%
As for NetSMF, we enumerate its edge samples $M$ from $\{1Tm, 2Tm, 4Tm, 8Tm\}$\footnote{We can not run the experiment with $M=10Tm$ in NetSMF paper because it needs 1.7TB memory but our machine has only 1.5TB memory.}.
Both \lightne{}, \lightneplus{} and NetSMF set $T=10$.
The predictive performance of \lightne{}, \lightneplus{}, NetSMF and ProNE+ is shown in Table~\ref{tbl:oag} and \Figref{fig:oag}.

\begin{table}[tb]
    \caption{\revise{Comparison results of node classification on OAG with label ratio 0.001\%, 0.01\%, 0.1\% and 1\%. 
    }}
    \label{tbl:oag}
    \centering
    \footnotesize
    \begin{tabular}{@{~}l@{~}|c|@{~}r@{~}|r@{~~}r@{~~}r@{~~}r@{~}}
        \toprule
                    Metric   & Method              & Time     & 0.001\%        & 0.01\%         & 0.1\%          & 1\%            \\\midrule
       \multirow{8}{*}{Micro}   & NetSMF (M=8Tm)      & 22.4~h   & 30.43          & 31.66          & 35.77          & 38.88          \\
                               & ProNE+               & 21~min   & 23.56          & 29.32          & 31.17          & 31.46          \\ \cmidrule{2-7}
                               & \lightne{}~(S)    & 20.9~min & 23.89          & 30.23          & 32.16          & 32.35          \\
                               & \lightneplus{}~(S)    & 11.8~min & 43.93          & 51.22         & 53.02          & 53.23          \\
                               & \lightne{}~(L)   & 1.53~h   & 44.50 & 52.89 & 54.98 & 55.23 \\
                               & \lightneplus{}~(L)    & 2.50~h   & \textbf{54.77} & \textbf{61.17} & \textbf{62.69} & \textbf{62.84} \\
                               \midrule
        \multirow{8}{*}{Macro} & NetSMF (M=8Tm)      & 22.4~h   & 7.84           & 9.34           & 13.72          & 17.82          \\
                               & ProNE+               & 21~min   & 10.47          & 10.30          & 9.83           & 9.79           \\\cmidrule{2-7}
                               & \lightne{}~(S)    & 20.9~min & 10.90          & 11.92          & 11.59          & 11.57          \\
                               & \lightneplus{}~(S)    & 11.8~min & 24.77          & 32.11          & 33.97         & 34.02         \\
                               & \lightne{}~(L)    & 1.53~h   & 25.85 & 35.72 & 38.18 & 38.53 \\ 
                               & \lightneplus{}~(L)    & 2.50~h   & \textbf{36.52} & \textbf{45.15} & \textbf{46.80} & \textbf{47.17} \\
        \bottomrule
    \end{tabular}.
    \normalsize
\end{table}

\diff{
\vpara{Comparing \lightneplus{}~(L) with NetSMF}
As shown in Table~\ref{tbl:oag}, \lightneplus{}~(L) achieves 9.0\(\times\) speedup~(2.50h v.s. 22.4h) 
and significant performance gain~(on average 77.5\% and 288.8\% relatively better regarding Micro and Macro F1, respectively). 

\vpara{Comparing \lightneplus{}~(S) with ProNE+}
As shown in Table~\ref{tbl:oag}, not only does \lightneplus{}~(S) run faster than ProNE+~(11.8~min v.s. 21~min),
but also outperforms ProNE+ significantly~(averagely +22.5 Micro F1 and +21.9 Macro F1). 

\vpara{Comparing \lightneplus{} with \lightne{}}
As shown in Table~\ref{tbl:oag}, 
\lightneplus{} achieves significantly better performance than \lightne{} regardless of the number of edge samples.
Regarding Micro and Macro F1, \lightneplus{}~(L) achieve averagely 16.6\% and 28.2\% relative performance gain than \lightne{}, and \lightneplus{}~(S)  outperforms \lightne{}~(S) by  70.7\% and 170.7\% on average.
The significant performance improvement of  \lightneplus{} illustrates the effectiveness of power iteration scheme (recall that we set $q=5$ for \lightneplus{} on OAG dataset). However, there is no free lunch. As we can see, \lightneplus{} (L)  spends more time than \lightne{} (L), mainly due to the extra computation caused by power iteration.
}


\emph{
    Overall, \lightneplus{} is a Pareto-optimal solution that strictly dominates ProNE+ or NetSMF --- for either ProNE+ or NetSMF, one can find a configuration of \lightneplus{} in Table~\ref{tbl:oag} 
    that is faster and more accurate even compared to \lightne{}~\cite{qiu2021lightne}.
    Moreover, there is a clear trade-off between
    efficiency and effectiveness of \lightneplus{} in \Figref{fig:oag} and \lightneplus{} has stronger scalability and better performance than \lightne{} thanks to power iteration scheme.
    That means a user can configure and adjust \lightneplus{} flexibly according to his/her time/cost budgets and performance requirements.
    Comparing \lightneplus{} to NetSMF and ProNE+ also suggests that
spectral propagation plays the role of ``standing on the shoulder of giants'' --- the quality of the enhanced embedding heavily relies on that of the initial one.
}

\subsubsection{Ablation Study on the OAG Dataset}
Next, we conduct ablation studies on the OAG dataset.
\begin{figure}[tb]
    \centering
    \includegraphics[width=0.5\textwidth]{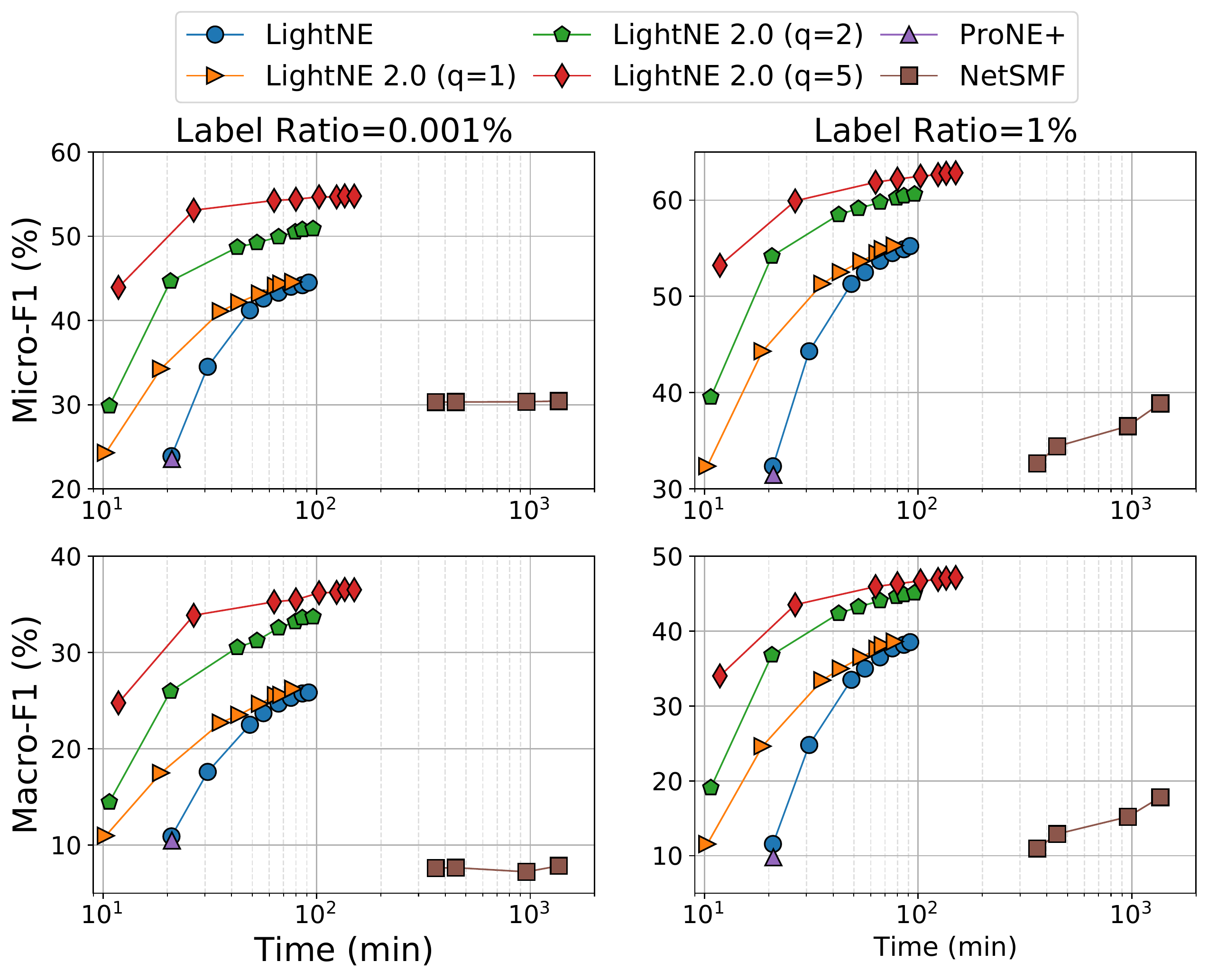}
    \caption{
        Efficiency-effectiveness trade-off curve of \lightneplus{}.
    }
    \label{fig:oag}
\end{figure}

\vpara{Ablation Study on Running Time}
We break down the running time of \lightne{}, \lightneplus{},  NetSMF,  and ProNE+, as shown in Table~\ref{tbl:time_dis}.
\lightne{} and \lightneplus{} consists of three stages --- parallel sparsifier
construction, randomized SVD, and spectral propagation.
In contrast, (1) NetSMF does not have the third stage; (2) ProNE+ directly factorizes a simple graph Laplacian matrix, so it doesn't have the first stage; \diff{(3) \lightneplus{} is enhanced by fast randomized SVD, which is faster and more scalable  than the vanilla randomized SVD in \lightne{}.

Comparing \lightneplus{} with NetSMF, \lightneplus{} achieves 33x speedup when constructing the sparsifier,
showing the advantages of the sparse parallel graph processing, the downsampling algorithm, and the sparse parallel hashing.
\lightneplus{}~(L, $q=1$) also achieves 6.2x speedup when factorizing the sparse matrix, showing the advantage of proposed fast randomized SVD algorithm and the advantage of Intel MKL over Eigen3 in the implementation of fast randomized SVD.

Comparing \lightneplus{} with ProNE+, 
\lightneplus{} use eigSVD as the economic SVD for the last stage of spectral propagation to achieve 6.1~min, while ProNE+ cost 8.1~min.
The factorization step in \lightneplus{} is significantly faster than that of ProNE+.
The one reason is that the matrix factorized by \lightneplus{}~(S) is more sparser than ProNE+ which makes the computation of SPMM cheaper. 
Note that ProNE+ has exactly $m$ non-zeros in its matrix to be factorized.
However, for \lightneplus{}~(S) with $M=0.077Tm=0.77m$ samples, the actual number of non-zeros in the sparsifier
could be fewer than $m$,
due to the downsampled PathSampling algorithm~(Algo.~\ref{alg:geomsample}). The other reason is that the application of fast orthonormalization operation further accelerate the computation of fast randomized SVD.

Comparing \lightneplus{} and \lightne{}, we have two observations.
First, the running time distribution of \lightneplus{} and \lightne{} mainly differs on randomized SVD, because \lightneplus{} replaces \lightne{}'s vanilla randomized SVD with fast randomized SVD while keeping the sparsifier construction and spectral propagation  the same.
Second, as the increase of the number of power iterations $q$, the fast randomized SVD of \lightneplus{} becomes more and more expensive. Especially, \lightneplus{}~(L) with $q=5$ is even slower than \lightne{}, indicating that the effectiveness of power iteration scheme of \lightneplus{} is not a free lunch.
\hide{Based on the above analysis,
we can explain the results in Table~\ref{tbl:time_dis}.
\lightneplus{}~(S, $q=1$) achieves 3.8x speedup using the proposed fast randomized SVD compared to \lightne{}~(S). 
However, \lightneplus{}~(L, $q=1$) is only slightly faster than \lightne{}~(L) with the acceleration of fast orthonormalization operation, because SPMM operation is the main computation part of power iteration for large edge samples. Combining Table~\ref{tbl:oag} with Table~\ref{tbl:time_dis}, it proves that power iteration will be a great enhanced technique especially for small edge samples. 
}
\begin{table}[tb]
    \caption{\revise{The running time distribution of  \lightneplus{}, \lightne{}, NetSMF and ProNE+ on OAG dataset. NA means the algorithm does not have the corresponding stage.}}
    \label{tbl:time_dis}
    \centering
    \footnotesize
    \begin{tabular}{@{~}c@{~}|@{}c@{}|@{}c@{}|@{}c@{}}
        \toprule
        Time& \multicolumn{1}{@{~}l@{~}|}{\begin{tabular}[c]{@{}c@{}}Parallel Sparsifier\\Construction\end{tabular}} & \multicolumn{1}{@{~}l@{~}|}{\begin{tabular}[c]{@{}c@{}}Randomized\\SVD \end{tabular}} & \multicolumn{1}{@{~}l@{~}}{\begin{tabular}[c]{@{}c@{}} Spectral\\Propagation\end{tabular}} \\\midrule
        \lightne{}~(L) & 32.8~min   & 49.9~min   & 8.1~min. \\
        \lightneplus{}~(L, $q=1$) &32.8~min & 38.8~min &6.1~min \\
        \lightneplus{}~(L, $q=5$) &32.8~min & 113.3~min &6.1~min \\
        NetSMF~(M=8Tm)   & 18~h   & 4~h  & NA            \\
        \midrule
        \lightne{}~(S) & 1.4~min & 10.5~min  & 8.1~min        \\
        \lightneplus{}~(S, $q=1$) & 1.4~min & 2.8~min  & 6.1~min        \\
        \lightneplus{}~(S, $q=5$) & 1.4~min & 4.3~min  & 6.1~min        \\
        ProNE+            & NA   & 12.0~min  & 8.1~min          \\
        \bottomrule
    \end{tabular}
    \normalsize
\end{table}

\vpara{Ablation Study on Power Iteration}
Before conducting the ablation study, we analyze the complexity of power iteration in terms of the number of edge samples. 
The power iteration scheme consists of two parts, sparse matrix-matrix multiplication~(SPMM) and the proposed fast orthonormalization (via eigSVD). 
As for SPMM, when choosing more edge samples, sparsifier construction step generates a sparse matrix with more non-zeros, making SPMM operation more expensive. 
As for fast orthonormalization, it takes the dense output matrix of SPMM as input, thus is irrelevant to the number of edge samples.

To empirically understand the impact of different times of power iteration~(the number of $q$) on performance of \lightneplus{}, we enumerate $q$ from $\{1,2,5\}$ and give a detailed ablation study on the $q$ parameter in \Figref{fig:oag}. Each curve in \Figref{fig:oag} is to fix all parameters except the edge samples, and then enumerates the edge samples from $\{0.077Tm,1Tm,5Tm,7Tm,10Tm,13Tm,15Tm,17Tm\}$ to get different runtime. 
First, we can observe that \lightneplus{}~($q\!=\!1$) shows almost the same Micro F1 and Macro F1 as \lightne{}~\cite{qiu2021lightne}. At the same time, we can observe that \lightneplus{}~($q\!=\!1$) is significant faster than \lightne{} especially for less edge samples, which proves the effect of fast orthonormalization operation in power iteration. 
As the increase of edge samples, the SPMM operation occupies the main computational overhead of power iteration, making the acceleration of fast orthonormalization operation less remarkable. \lightneplus{}~($q\!=\!2$) show significantly better performance than \lightneplus{}~($q\!=\!1$), and \lightneplus{}~($q=5$) is even better. This proves that the power iteration steadily improve the performance at the cost of slightly increased runtime. 
}

\vpara{Ablation Study on Sample Size} 
Comparing to NetSMF with $8Tm$ samples, \lightneplus{}~(L) is able to draw up to $17Tm$~(2\(\times\)) samples and achieves significantly better performance. 
The large sample size can be attributed to (1) compressed GBBS, (2) downsampling technique, and (3) sparse parallel hashing.
However, the uncompressed OAG graph~(in CSR format) occupies
only 16GB; thus, the effect to compressed GBBS is 
negligible given our machine has 1.5TB memory~(though compressed GBBS plays a big role in very large graphs, ref. Section~\ref{subsec:verylarge}). 
We turn off the downsampling in \lightneplus{} and gradually increase its number of samples until out-of-memory.
We observe that we can have at most $12.5Tm$ samples without downsampling, which is 56.3\% greater than NetSMF's $8Tm$.
The above analysis suggests that the sparse parallel hashing is another contributor to the larger sample size --- it increases affordable sample size by 56.3\%, and downsampling further increases by 60\%.
This is because the shared-memory hash table in \lightneplus{} can significantly save memory. However, NetSMF maintains a thread-local sparsifer in each thread and merges them at the end of sampling.

\hide{
\vpara{Effect of Spectral Propagation} We finally investigate the
impact of spectral propagation. We report the results of \lightne{}-small
and \lightne{}-Large
without the spectral propagation in Table~\ref{tbl:oag}, 
denoted by \lightne{}-Small~(w/o SP) and \lightne{}-Large~(w/o SP), respectively.
Comparing \lightne{}-Large and \lightne{}-Large~(w/o SP), 
we can see that spectral propagation significantly boosts the performance of the initial NetSMF embedding.
Comparing \lightne{}-Small and \lightne{}-Small~(w/o SP), it is interesting that spectral propagation
improves Macro F1 at a slight cost of Micro F1.   
Moreover, comparing \lightne{}-Large~(with $M=20Tm$ samples) and NetSMF~(with $M=8Tm$ samples),
we can see how the increase of edge samples improves the embedding quality, 
which is impossible without our careful co-design of algorithm and system.

Overall, the results suggest that
spectral propagation plays the role of ``standing on the shoulder of giants'' --- the quality of the enhanced embedding heavily relies on that of the initial one.
}

\subsection{Very Large Graphs}
\label{subsec:verylarge}
\hide{
    \begin{table}[htbp]
        \caption{Large graph statistics.}
        \label{tbl:stats}
        \centering
        \small
        \begin{tabular}{rrrr}
            \toprule
                            & ClueWeb        & Hyperlink2014   & Hyperlink2012   \\\midrule
            $|V|$           & 978,408,098    & 1,724,573,718   & 3,563,602,789   \\
            $|E|$           & 74,744,358,622 & 124,141,874,032 & 225,840,663,232 \\
            CSR memory~(GB) & 564            & 938             & 1709            \\
            Compressed~(GB) & 107            & 197             & 371             \\
            \bottomrule
        \end{tabular}
        \normalsize
    \end{table}
}

To further illustrate the lightweightness and scalability of \lightneplus{},
we test \lightneplus{} on two very large 100-billion scale graphs,
ClueWeb-Sym, and Hyperlink2014-Sym, as shown in Table~\ref{tbl:data}.
It is worth noting that none of the existing network embedding systems can handle such large graphs in a single machine.
For example, it takes 564GB memory to store ClueWeb-Sym's 74 billion unweighted edges.
Furthermore, the Hyperlink2014 graph is one of the largest publicly available graph today, and very few graph processing systems or graph algorithms have been applied to a graph of this magnitude, in any setting~\cite{dhulipala2018theoretically}.
However, by adopting the graph compression from GBBS~\cite{dhulipala2018theoretically},
we are able to reduce the size of ClueWeb-Sym to 107GB.
Moreover, by leveraging the downsampling technique in \Secref{subsec:algorithm},
we are able to maintain a $O(n\log{n})$ sparsifier, which only requires a modest amount of memory and enables us to apply fast randomized SVD without an excessive memory footprint.

To evaluate the performance of \lightneplus{} on very large graphs,
we adopt link prediction to be the evaluation task, as
vertex labels are not available for these graphs.
We follow \shortpbg{} to set up link prediction evaluation ---  
we randomly exclude 0.00001\%  edges from the training graph for evaluation.
When training \lightneplus{} on the two very large graphs,
we skip the spectral propagation step~(due to memory issue) and set $T=2$ as well as $d=32$.
After training,
the ranking metrics on the test set
are obtained by ranking positive edges among randomly sampled corrupted edges.
We gradually increase the number of edge samples until it reaches the 1.5TB memory bottleneck.
\begin{figure}[t]
    \centering
    \includegraphics[width=0.5\textwidth]{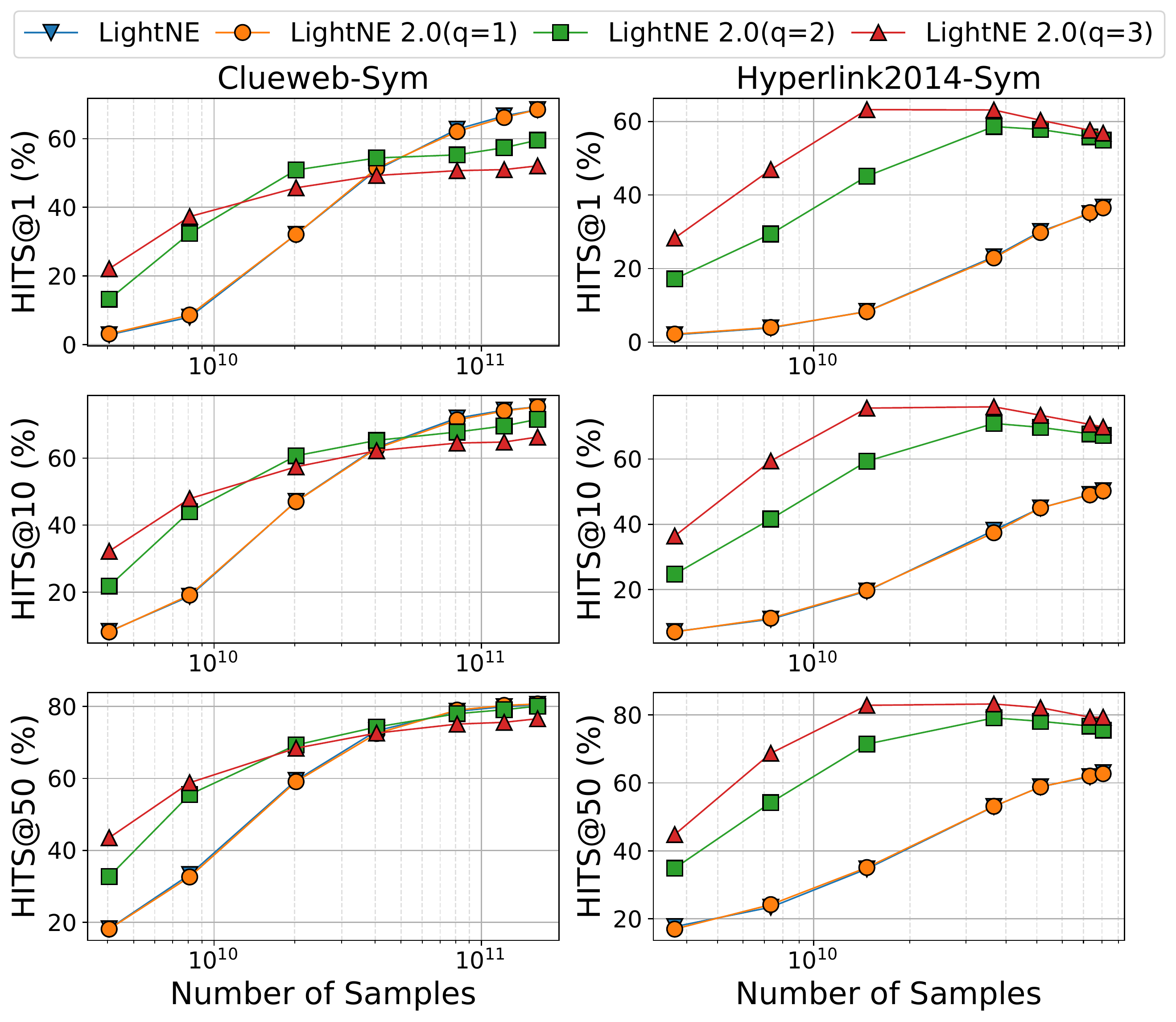}
    \caption{
        HITS@K~($K=1,10,50$) of \lightneplus{} and \lightne{} w.r.t. the number of samples.
    }
    \label{fig:large}
\end{figure}
\Figref{fig:large} presents the HITS@K with $K=1,10,50$ of \lightneplus{} with different numbers of edge samples $M$.
As we can see, the more samples we draw by edge sampling, the higher accuracy \lightneplus{} can achieve.
Moreover, the trend of growth shown in \Figref{fig:large}
suggests that the performance in these datasets can be further improved if we can
overcome the memory bottleneck by, for example, using a machine with larger memory, or designing compressed hash tables and linear algebra tools.

\diff{
Comparing \lightneplus{} with \lightne{}~\cite{qiu2021lightne}, we can observe that \lightne{}'s performance is largely improved by power iteration ($q=2$ or $q=3$) especially for less edge samples. However, it is worth noting that on  Clueweb-Sym dataset with more edge samples, more power iteration is less effective. Therefore, hyperparameter $q$ should be carefully selected when having more edge samples.
\begin{figure}[t]
    \centering
    \includegraphics[width=0.5\textwidth]{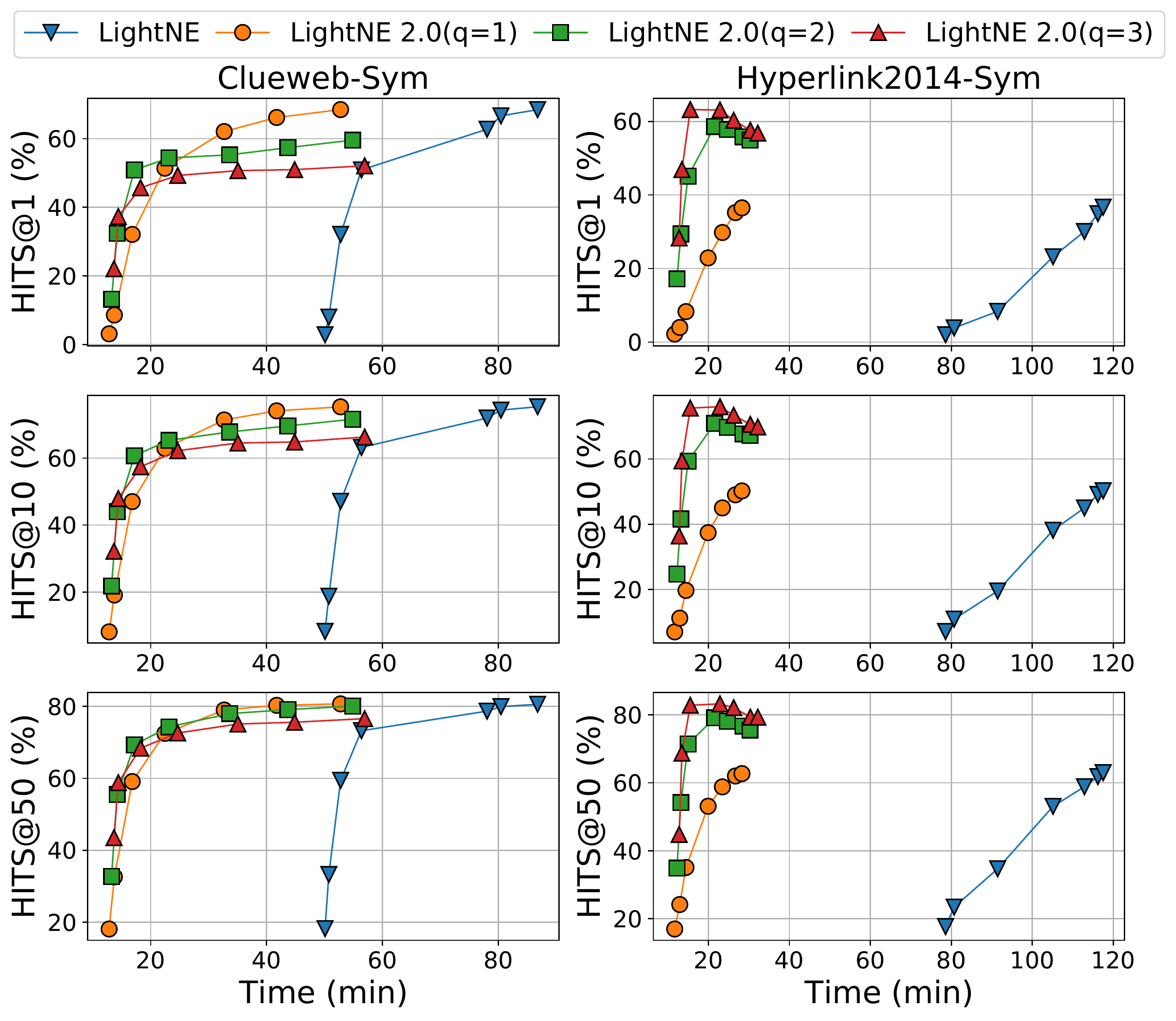}
    \caption{
    \revise{Efficiency-effectiveness trade-off curve of \lightneplus{} with different $q$ for very large graphs (run with 88 virtual threads).}
    }
    \label{fig:large_trade_off}
\end{figure}

As for efficiency, we have the following two observations from \Figref{fig:large_trade_off}. First, \lightneplus{} is much faster than \lightne{}. With less edge samples,
\lightneplus{}~($q=1$) is able to embed very large graphs in half an hour, achieving about 5x speedup on Clueweb-Sym and 6x speedup comparing to \lightne{}.
Second, comparing \lightneplus{}~($q\!=\!1,2,3$)  shows that each power iteration requires (nearly) constant and little running time.
%
Furthermore, the efficiency of \lightneplus{} allows automated  hyperparameter tuning to improve performance.
}

\hide{
We also report the training time and the SVD time  of \lightne{} in
the following table:
    \begin{center}
        \small
        \begin{tabular}{@{~}l|rrrrr@{~}}
            \toprule
            Dataset           & Time  & SVD Time & MR     & MRR  & Hits@10 \\\midrule
            ClueWeb-Sym       & 1.1~h & 0.79~h   & 117.37 & 0.59 & 0.66    \\
            Hyperlink2014-Sym & 1.9~h & 1.4~h    & 148.88 & 0.40 & 0.49    \\
            \bottomrule
        \end{tabular}.
        \normalsize
    \end{center}
}

\hide{
    hyperlink2012 6.27e+09 samples
}

\diff{
\subsection{Automated Hyperparameter Tuning with FLAML}
In this section, we incorporate automated hyperparameter tuning into \lightneplus{}. In particular, we use a state-of-the-art AutoML library FLAML to search hyperparemeters, and refer \lightneplus{}~(FLAML) to the automatically tuned \lightneplus{}.
We run the tuning process with up to 20 iterations and use the metric in downstream tasks (Micro/Macro-F1, HITS@$K$) as the searching objective.
We report the results of \lightneplus{}~(FLAML) on OAG in Table~\ref{tbl:tune},  and the results on ClueWeb-Sym and Hyperlink2014-Sym in Table~\ref{tbl:tunelarge}. \lightneplus{} is the best result in the experiment above without tuning hyperparameters. As we can see, tuning hyperparameters with FLAML consistently improves the performance of \lightneplus{}.

\hide{
 The results of \lightneplus{} on OAG when choosing small edge samples is improved by 1.0 in Micro F1 and 0.75 in Macro F1 on average. 
 In OAG dataset, it only takes 16 iteration for FLAML to get the improved result and additionally cost 7.2 hours. 
In Table~\ref{tbl:tunelarge}, ClueWeb-Sym is significantly improved with only 7 iteration. 
Hyperlink2014-Sym is significantly improved with only 16 iteration. 
}
 
\vpara{Case Study on BlogCatalog}
We run the tuning process 50 iterations and use the average Macro F1 as the objective for BlogCatalog. 
 \Figref{fig:tune}~(a) shows the fast convergence of hyperparameters tuning where the red line indicates the best average Macro F1 during the tuning process. We 
can see that FLAML converges very quickly.
We then analyze the effects of tuned coefficients for random walk matrix polynomials.
As shown in \Figref{fig:tune}~(b), the blue bars indicate the histogram of eigenvalues of the matrix $\frac{1}{T}\sum_{r=1}^T (\bm{D}^{-1}\bm{A})^r \bm{D}^{-1}$ and the red bars indicate the histogram of eigenvalues of the matrix $\sum_{r=1}^T s_r(\bm{D}^{-1}\bm{A})^r \bm{D}^{-1}$. It shows that the tuned random walk matrix polynomial acts as a band-pass filter for eigenvalues, keeping eigenvalues relevant to downstream tasks.

\begin{table}[t]
    \caption{\revise{Node classification results on OAG with tuned parameters.}}
    \label{tbl:tune}
    \centering
    \small
    \begin{tabular}{@{~}l@{~}|@{~}c@{~}c@{~}c@{~}c@{~}c@{~}}
        \toprule
                         Metric          & Label Ratio~(\%) & 0.001\%              & 0.01\%              & 0.1\%       & 1\%  \\\midrule
      \multirow{2}{*}{Micro F1}  & \lightneplus{}   & 54.77         & 61.17         & 62.69          & 62.84          \\
                     & \lightneplus{}~(FLAML)     & \textbf{55.23}          & \textbf{61.64}          & \textbf{63.04}          & \textbf{63.19}          \\
        \midrule
                 \multirow{2}{*}{Macro F1} & \lightneplus{}   & 36.52        & 45.15        & 46.80         & 47.17          \\
                   & \lightneplus{}~(FLAML)     & \textbf{37.63}          & \textbf{45.33}          & \textbf{47.26}          & \textbf{47.58}          \\
        \bottomrule
    \end{tabular}.
\end{table}

\begin{table}[t]
    \caption{\revise{Link prediction results on ClueWeb-Sym and Hyperlink2014-Sym with tuned parameters. CS and HS stand for ClueWeb-Sym and Hyperlink2014-Sym, respectively.}}
    \label{tbl:tunelarge}
    \centering
    \small
    \begin{tabular}{@{~}l@{~}|@{~}c@{~}c@{~}c@{~}c@{~}c@{~}}
        \toprule
        Dataset          & Metric  & HITS@1    & HITS@10      & HITS@50   \\\midrule                          
        \multirow{2}{*}{CS} & \lightneplus{}  & 0.680          & 0.750       & 0.800     \\
             &    \lightneplus{}~(FLAML)   & \textbf{0.705} & \textbf{0.766} & \textbf{0.814}  \\\midrule
        \multirow{2}{*}{HS}  &  \lightneplus{}  & 0.631          & 0.754       & 0.826      \\
             &    \lightneplus{}~(FLAML)   & \textbf{0.687} & \textbf{0.808} & \textbf{0.870}  \\
        \bottomrule
    \end{tabular}
\end{table}

\begin{figure}[t]
     \centering
     \subfloat[Tuning process for \lightneplus{} on BlogCatalog.]{
         \includegraphics[width=0.24\textwidth]{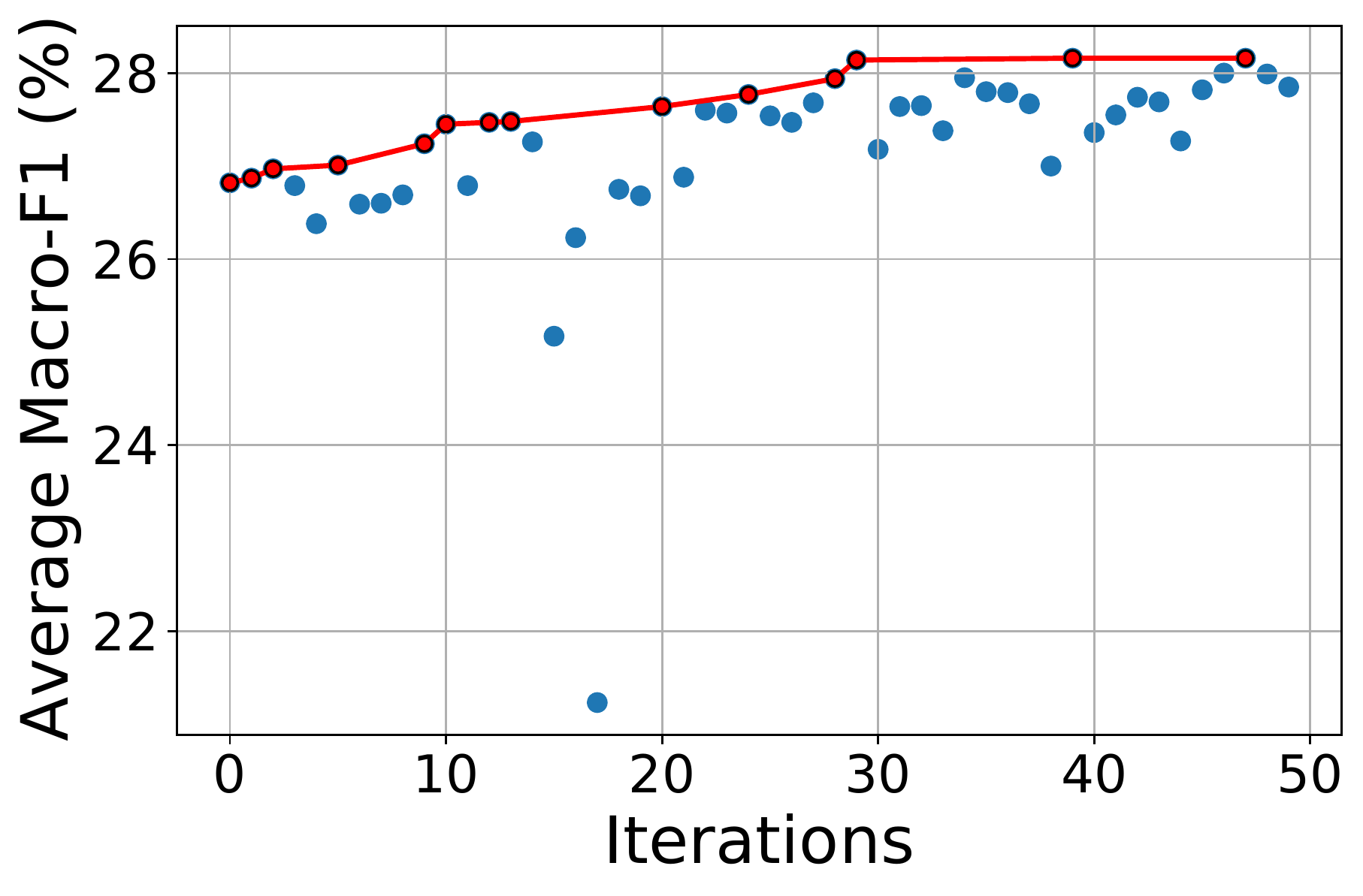}
     }
     \subfloat[Histogram of eigenvalues of random walk matrix polynomials on BlogCatalog.]{
         \includegraphics[width=0.25\textwidth]{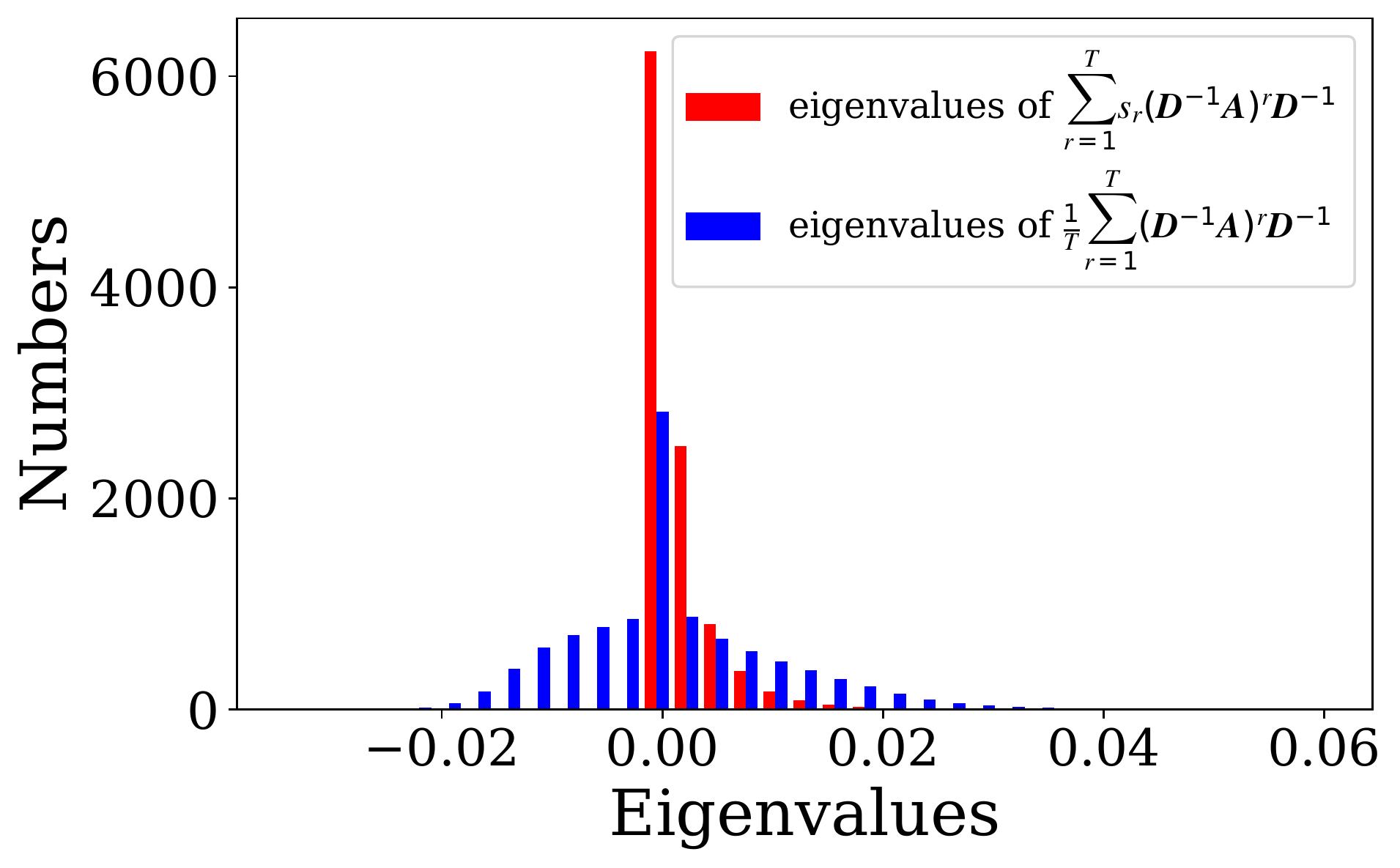}
     }
     \caption{
        The effects of tuning parameters for \lightneplus{} with FLAML.
     }
     \label{fig:tune}
\end{figure}

\subsection{Scalability}
We study the scalability of \lightneplus{} by analyzing the relationship between wall-clock time and number of threads. We use a single-machine shared memory implementation with multi-threading acceleration. We report the running time of \lightneplus{} when setting the number of
threads to be 1, 3, 5, 10, 20, 40, 60, 88 respectively. As shown in \Figref{fig:threads}, \lightneplus{} takes 15.8 hours to embed the Hyperlink2014 network with one thread and 47 mins to run with 40 threads, achieving a 20 speedup ratio (with ideal being 40). This relatively good sub-linear speedup supports \lightneplus{} to scale up to very large graphs.
\begin{figure}[t]
    \centering
    \includegraphics[width=0.3\textwidth]{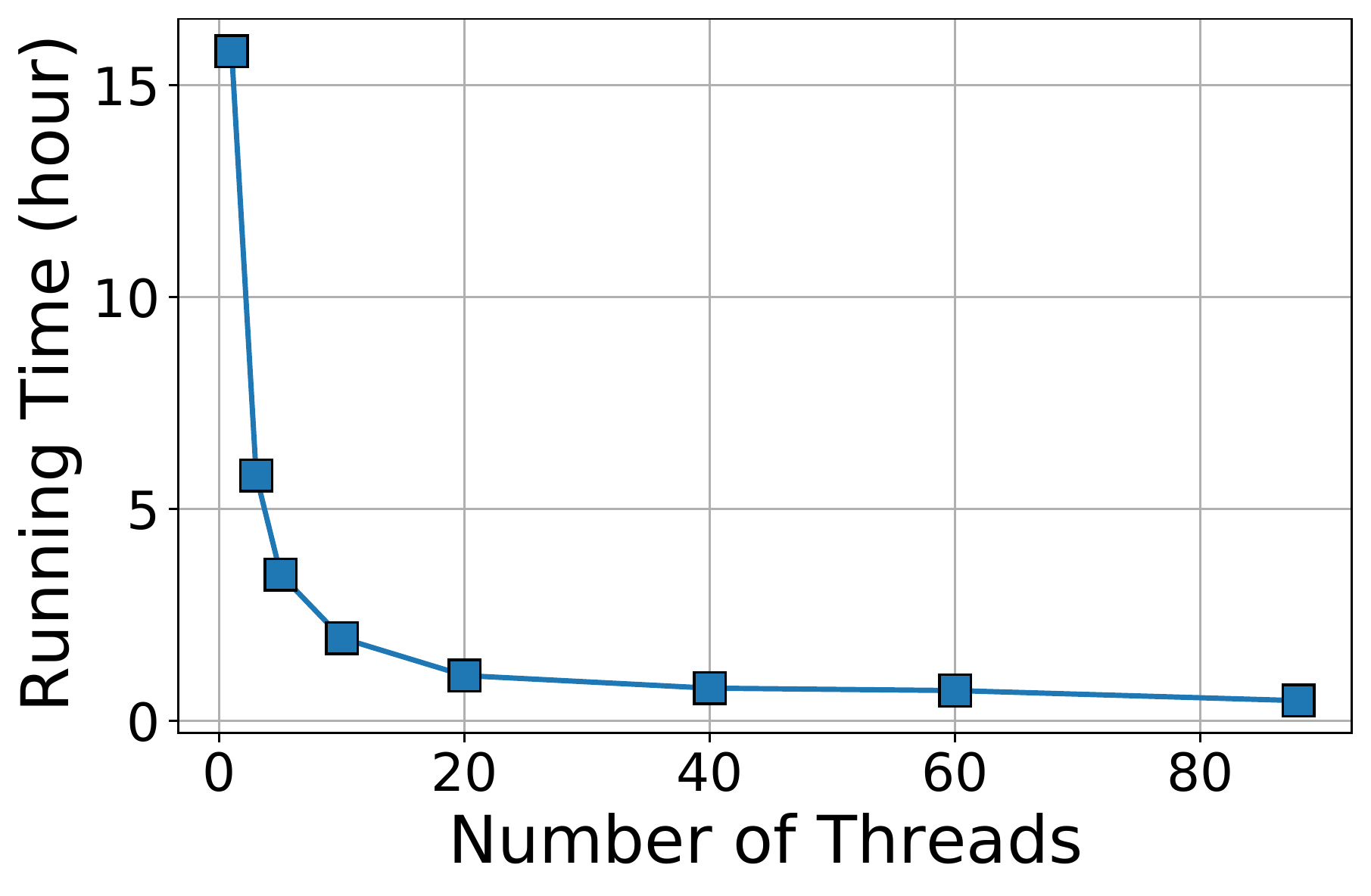}
    \caption{
    \revise{Runtime of \lightneplus{}~($q$=3) v.s. the number of threads, on Hyperlink2014. The number of samples is 8.1$\times$ $10^{10}$.}
    }
    \label{fig:threads}
\end{figure}
}


%% file: 6.conclusion.tex
\section{Conclusion}
\label{sec:conclusion}

In this work, we present \lightneplus{},  a single-machine shared-memory system that significantly improves the efficiency, scalability, and accuracy of state-of-the-art network embedding techniques.
\lightneplus{} combines two advanced network embedding algorithms, NetSMF and ProNE, to achieve state-of-the-art performance on nine benchmarking graph datasets, compared to four recent network embedding systems---\gvt{}, \pbg{}, NetSMF, and \lightne{}.
By incorporating sparse parallel graph processing techniques, and other parallel algorithmic techniques like sparse parallel hashing, high-performance parallel linear algebra, \diff{and fast randomized SVD, \lightneplus{} is able to learn high-quality embeddings for graphs with hundreds of billions of edges in half an hour, all at a modest cost. 
Moreover, the efficiency of \lightneplus{} allows us to further improve embedding quality by  automated hyperparemeter tuning.}

%% file: bare_jrnl_compsoc.bbl
\begin{thebibliography}{10}
\providecommand{\url}[1]{#1}
\csname url@samestyle\endcsname
\providecommand{\newblock}{\relax}
\providecommand{\bibinfo}[2]{#2}
\providecommand{\BIBentrySTDinterwordspacing}{\spaceskip=0pt\relax}
\providecommand{\BIBentryALTinterwordstretchfactor}{4}
\providecommand{\BIBentryALTinterwordspacing}{\spaceskip=\fontdimen2\font plus
\BIBentryALTinterwordstretchfactor\fontdimen3\font minus
  \fontdimen4\font\relax}
\providecommand{\BIBforeignlanguage}[2]{{%
\expandafter\ifx\csname l@#1\endcsname\relax
\typeout{** WARNING: IEEEtran.bst: No hyphenation pattern has been}%
\typeout{** loaded for the language `#1'. Using the pattern for}%
\typeout{** the default language instead.}%
\else
\language=\csname l@#1\endcsname
\fi
#2}}
\providecommand{\BIBdecl}{\relax}
\BIBdecl

\bibitem{ying2018graph}
R.~Ying, R.~He, K.~Chen, P.~Eksombatchai, W.~L. Hamilton, and J.~Leskovec,
  ``Graph convolutional neural networks for web-scale recommender systems,'' in
  \emph{KDD '18}, 2018, pp. 974--983.

\bibitem{wang2018billion}
J.~Wang, P.~Huang, H.~Zhao, Z.~Zhang, B.~Zhao, and D.~L. Lee, ``Billion-scale
  commodity embedding for e-commerce recommendation in alibaba,'' in \emph{KDD
  '18}, 2018, pp. 839--848.

\bibitem{ramanath2018towards}
R.~Ramanath, H.~Inan, G.~Polatkan, B.~Hu, Q.~Guo, C.~Ozcaglar, X.~Wu,
  K.~Kenthapadi, and S.~C. Geyik, ``Towards deep and representation learning
  for talent search at linkedin,'' in \emph{CIKM '18}, 2018, pp. 2253--2261.

\bibitem{tang2015line}
J.~Tang, M.~Qu, M.~Wang, M.~Zhang, J.~Yan, and Q.~Mei, ``Line: Large-scale
  information network embedding,'' in \emph{WWW '15}, 2015, pp. 1067--1077.

\bibitem{perozzi14deepwalk}
B.~Perozzi, R.~Al-Rfou, and S.~Skiena, ``Deepwalk: Online learning of social
  representations,'' in \emph{KDD '14}.\hskip 1em plus 0.5em minus 0.4em\relax
  ACM, 2014, pp. 701--710.

\bibitem{grover2016node2vec}
A.~Grover and J.~Leskovec, ``node2vec: Scalable feature learning for
  networks,'' in \emph{KDD '16}, 2016, pp. 855--864.

\bibitem{zhu2019graphvite}
Z.~Zhu, S.~Xu, J.~Tang, and M.~Qu, ``Graphvite: A high-performance cpu-gpu
  hybrid system for node embedding,'' in \emph{The World Wide Web
  Conference}.\hskip 1em plus 0.5em minus 0.4em\relax ACM, 2019, pp.
  2494--2504.

\bibitem{qiu2019netsmf}
J.~Qiu, Y.~Dong, H.~Ma, J.~Li, C.~Wang, K.~Wang, and J.~Tang, ``{NetSMF}:
  Large-scale network embedding as sparse matrix factorization,'' in
  \emph{WWW'19}, 2019, pp. 1509--1520.

\bibitem{zhang2019prone}
J.~Zhang, Y.~Dong, Y.~Wang, J.~Tang, and M.~Ding, ``Prone: fast and scalable
  network representation learning,'' in \emph{IJCAI '19}, 2019, pp. 4278--4284.

\bibitem{qiu2021lightne}
J.~Qiu, L.~Dhulipala, J.~Tang, R.~Peng, and C.~Wang, ``Lightne: A lightweight
  graph processing system for network embedding,'' in \emph{SIGMOD '21}, 2021,
  pp. 2281--2289.

\bibitem{wang2021flaml}
C.~Wang, Q.~Wu, M.~Weimer, and E.~Zhu, ``Flaml: A fast and lightweight automl
  library,'' in \emph{MLSys '21}, 2021.

\bibitem{hamilton2017representation}
W.~L. {Hamilton}, R.~{Ying}, and J.~{Leskovec}, ``Representation learning on
  graphs: Methods and applications.'' \emph{IEEE Data(base) Engineering
  Bulletin}, vol.~40, pp. 52--74, 2017.

\bibitem{mikolov2013distributed}
T.~Mikolov, I.~Sutskever, K.~Chen, G.~S. Corrado, and J.~Dean, ``Distributed
  representations of words and phrases and their compositionality,'' in
  \emph{Advances in neural information processing systems}, 2013, pp.
  3111--3119.

\bibitem{de2014global}
C.~De~Sa, K.~Olukotun, and C.~R{\'e}, ``Global convergence of stochastic
  gradient descent for some non-convex matrix problems,'' \emph{arXiv preprint
  arXiv:1411.1134}, 2014.

\bibitem{eckart1936approximation}
C.~Eckart and G.~Young, ``The approximation of one matrix by another of lower
  rank,'' \emph{Psychometrika}, vol.~1, no.~3, pp. 211--218, 1936.

\bibitem{cao2015grarep}
S.~Cao, W.~Lu, and Q.~Xu, ``Grarep: Learning graph representations with global
  structural information,'' in \emph{CIKM '15}, 2015, pp. 891--900.

\bibitem{ou2016asymmetric}
M.~Ou, P.~Cui, J.~Pei, Z.~Zhang, and W.~Zhu, ``Asymmetric transitivity
  preserving graph embedding,'' in \emph{KDD '16}, 2016, pp. 1105--1114.

\bibitem{qiu2018network}
J.~Qiu, Y.~Dong, H.~Ma, J.~Li, K.~Wang, and J.~Tang, ``Network embedding as
  matrix factorization: Unifying deepwalk, line, pte, and node2vec,'' in
  \emph{WSDM '18}.\hskip 1em plus 0.5em minus 0.4em\relax ACM, 2018, pp.
  459--467.

\bibitem{zhang2018billion}
Z.~Zhang, P.~Cui, H.~Li, X.~Wang, and W.~Zhu, ``Billion-scale network embedding
  with iterative random projection,'' in \emph{ICDM '18}.\hskip 1em plus 0.5em
  minus 0.4em\relax IEEE, 2018, pp. 787--796.

\bibitem{chen2019fast}
H.~Chen, S.~F. Sultan, Y.~Tian, M.~Chen, and S.~Skiena, ``Fast and accurate
  network embeddings via very sparse random projection,'' in \emph{CIKM '19},
  2019, pp. 399--408.

\bibitem{battaglia2018relational}
P.~W. Battaglia, J.~B. Hamrick, V.~Bapst, A.~Sanchez-Gonzalez, V.~Zambaldi,
  M.~Malinowski, A.~Tacchetti, D.~Raposo, A.~Santoro, R.~Faulkner
  \emph{et~al.}, ``Relational inductive biases, deep learning, and graph
  networks,'' \emph{arXiv preprint arXiv:1806.01261}, 2018.

\bibitem{kipf2016semi}
T.~N. Kipf and M.~Welling, ``Semi-supervised classification with graph
  convolutional networks,'' 2017.

\bibitem{velivckovic2017graph}
P.~Veli{\v{c}}kovi{\'c}, G.~Cucurull, A.~Casanova, A.~Romero, P.~Lio, and
  Y.~Bengio, ``Graph attention networks,'' \emph{arXiv preprint
  arXiv:1710.10903}, 2017.

\bibitem{xu2018powerful}
K.~Xu, W.~Hu, J.~Leskovec, and S.~Jegelka, ``How powerful are graph neural
  networks?'' \emph{arXiv preprint arXiv:1810.00826}, 2018.

\bibitem{lerer2019pytorch}
A.~Lerer, L.~Wu, J.~Shen, T.~Lacroix, L.~Wehrstedt, A.~Bose, and
  A.~Peysakhovich, ``Pytorch-biggraph: A large-scale graph embedding system,''
  \emph{arXiv preprint arXiv:1903.12287}, 2019.

\bibitem{yang2020homogeneous}
R.~Yang, J.~Shi, X.~Xiao, Y.~Yang, and S.~S. Bhowmick, ``Homogeneous network
  embedding for massive graphs via reweighted personalized pagerank,''
  \emph{Proceedings of the VLDB Endowment}, vol.~13, no.~5, pp. 670--683, 2020.

\bibitem{cheng15sample}
D.~Cheng, Y.~Cheng, Y.~Liu, R.~Peng, and S.~Teng, ``Efficient sampling for
  {G}aussian graphical models via spectral sparsification,'' in \emph{The 28th
  Conference on Learning Theory}, 2015, pp. 364--390.

\bibitem{HalkoMT11}
N.~Halko, P.-G. Martinsson, and J.~A. Tropp, ``Finding structure with
  randomness: Probabilistic algorithms for constructing approximate matrix
  decompositions,'' \emph{SIAM review}, vol.~53, no.~2, pp. 217--288, 2011.

\bibitem{teng2016scalable}
S.-H. Teng, ``Scalable algorithms for data and network analysis,''
  \emph{Foundations and Trends{\textregistered} in Theoretical Computer
  Science}, vol.~12, no. 1--2, pp. 1--274, 2016.

\bibitem{spielman2011graph}
D.~A. Spielman and N.~Srivastava, ``Graph sparsification by effective
  resistances,'' \emph{SIAM Journal on Computing}, vol.~40, no.~6, pp.
  1913--1926, 2011.

\bibitem{koutis2012improved}
I.~Koutis, A.~Levin, and R.~Peng, ``Improved spectral sparsification and
  numerical algorithms for sdd matrices,'' in \emph{29th International
  Symposium on Theoretical Aspects of Computer Science}, 2012, p. 266.

\bibitem{lovasz1993random}
L.~Lov{\'a}sz \emph{et~al.}, ``Random walks on graphs: A survey,''
  \emph{Combinatorics, Paul erdos is eighty}, vol.~2, no.~1, pp. 1--46, 1993.

\bibitem{SpielmanT11}
D.~Spielman and S.~Teng, ``Spectral sparsification of graphs,'' \emph{SIAM
  Journal on Computing}, vol.~40, no.~4, pp. 981--1025, 2011.

\bibitem{tang2009relational}
L.~Tang and H.~Liu, ``Relational learning via latent social dimensions,'' in
  \emph{KDD '09}, 2009.

\bibitem{qiu2020matrix}
J.~Qiu, C.~Wang, B.~Liao, R.~Peng, and J.~Tang, ``A matrix chernoff bound for
  markov chains and its application to co-occurrence matrices,'' \emph{NeurIPS
  '20}, 2020.

\bibitem{mohaisen2010measuring}
A.~Mohaisen, A.~Yun, and Y.~Kim, ``Measuring the mixing time of social
  graphs,'' in \emph{Proceedings of the 10th ACM SIGCOMM conference on Internet
  measurement}, 2010, pp. 383--389.

\bibitem{randnla}
P.~Drineas and M.~W. Mahoney, ``Randnla: Randomized numerical linear algebra,''
  \emph{Commun. ACM}, vol.~59, no.~6, p. 80–90, May 2016.

\bibitem{musco2015nips}
C.~Musco and C.~Musco, ``Randomized block krylov methods for stronger and
  faster approximate singular value decomposition,'' in \emph{Proceedings of
  the 28th International Conference on Neural Information Processing Systems -
  Volume 1}, ser. NIPS'15, 2015, p. 1396–1404.

\bibitem{frpca}
X.~Feng, Y.~Xie, M.~Song, W.~Yu, and J.~Tang, ``Fast randomized pca for sparse
  data,'' in \emph{The 10th Asian Conference on Machine Learning}, 14--16 Nov
  2018, pp. 710--725.

\bibitem{li2006very}
P.~Li, T.~J. Hastie, and K.~W. Church, ``Very sparse random projections,'' in
  \emph{KDD '06}, 2006, pp. 287--296.

\bibitem{tropp2019streaming}
J.~A. Tropp, A.~Yurtsever, M.~Udell, and V.~Cevher, ``Streaming low-rank matrix
  approximation with an application to scientific simulation,'' \emph{SIAM
  Journal on Scientific Computing}, vol.~41, no.~4, pp. A2430--A2463, 2019.

\bibitem{wang2021blendsearch}
C.~Wang, Q.~Wu, S.~Huang, and A.~Saied, ``Economical hyperparameter
  optimization with blended search strategy,'' in \emph{ICLR'21}, 2021.

\bibitem{abu2018watch}
S.~Abu-El-Haija, B.~Perozzi, R.~Al-Rfou, and A.~A. Alemi, ``Watch your step:
  Learning node embeddings via graph attention,'' \emph{NeurIPS '19}, vol.~31,
  2018.

\bibitem{cheng2015spectral}
D.~Cheng, Y.~Cheng, Y.~Liu, R.~Peng, and S.-H. Teng, ``Spectral sparsification
  of random-walk matrix polynomials,'' \emph{arXiv preprint arXiv:1502.03496},
  2015.

\bibitem{hou2021automated}
Z.~Hou, Y.~Cen, Y.~Dong, J.~Zhang, and J.~Tang, ``Automated unsupervised graph
  representation learning,'' \emph{TKDE '21}, 2021.

\bibitem{dhulipala2018theoretically}
L.~Dhulipala, G.~E. Blelloch, and J.~Shun, ``Theoretically efficient parallel
  graph algorithms can be fast and scalable,'' in \emph{SPAA'18}, 2018, pp.
  393--404.

\bibitem{shun2013ligra}
J.~Shun and G.~E. Blelloch, ``Ligra: a lightweight graph processing framework
  for shared memory,'' in \emph{ACM Sigplan Notices}, vol.~48, no.~8.\hskip 1em
  plus 0.5em minus 0.4em\relax ACM, 2013, pp. 135--146.

\bibitem{kepner2011graph}
J.~Kepner and J.~Gilbert, \emph{Graph algorithms in the language of linear
  algebra}.\hskip 1em plus 0.5em minus 0.4em\relax SIAM, 2011.

\bibitem{shun2015ligraplus}
J.~Shun, L.~Dhulipala, and G.~E. Blelloch, ``Smaller and faster: Parallel
  processing of compressed graphs with {Ligra+},'' in \emph{DCC}, 2015.

\bibitem{semisort}
Y.~Gu, J.~Shun, Y.~Sun, and G.~E. Blelloch, ``A top-down parallel semisort,''
  in \emph{SPAA '15}, 2015, pp. 24--34.

\bibitem{milk}
V.~Kiriansky, Y.~Zhang, and S.~Amarasinghe, ``Optimizing indirect memory
  references with milk,'' in \emph{Proceedings of the 2016 International
  Conference on Parallel Architectures and Compilation}, 2016, pp. 299--312.

\bibitem{maier2016concurrent}
T.~Maier, P.~Sanders, and R.~Dementiev, ``Concurrent hash tables: Fast and
  general?(!),'' \emph{ACM SIGPLAN Notices}, vol.~51, no.~8, pp. 1--2, 2016.

\bibitem{shun13reducing}
J.~Shun, G.~E. Blelloch, J.~T. Fineman, and P.~B. Gibbons, ``Reducing
  contention through priority updates,'' in \emph{SPAA}, 2013.

\bibitem{fan2008liblinear}
R.-E. Fan, K.-W. Chang, C.-J. Hsieh, X.-R. Wang, and C.-J. Lin, ``Liblinear: A
  library for large linear classification,'' \emph{JMLR '08}, vol.~9, pp.
  1871--1874, 2008.

\bibitem{tsoumakas2009mining}
G.~Tsoumakas, I.~Katakis, and I.~Vlahavas, ``Mining multi-label data,'' in
  \emph{Data mining and knowledge discovery handbook}.\hskip 1em plus 0.5em
  minus 0.4em\relax Springer, 2009, pp. 667--685.

\bibitem{sinha2015overview}
A.~Sinha, Z.~Shen, Y.~Song, H.~Ma, D.~Eide, B.-J. Hsu, and K.~Wang, ``An
  overview of microsoft academic service (mas) and applications,'' in \emph{WWW
  '15}, 2015, pp. 243--246.

\end{thebibliography}
